\documentclass[12pt]{article}
\usepackage{amsmath,amsthm,amssymb,amsfonts,fullpage,url,soul,natbib,thmtools,listings}
\usepackage[dvipsnames]{xcolor}
\usepackage[normalem]{ulem}
\usepackage{booktabs,colortbl,cancel,multirow,titling}
\usepackage{algorithm,bbm,graphicx}
\usepackage[hidelinks]{hyperref}
\usepackage[]{algpseudocode}
\usepackage[titletoc,toc,title]{appendix}
\usepackage[T1]{fontenc}
\usepackage{pifont}
\usepackage{arydshln}

\def\lQ{\scalebox{-1}[1]{''}}

\newcommand{\mc}[2]{\multicolumn{#1}{c}{#2}}
\definecolor{Gray}{gray}{0.85}
\definecolor{LightCyan}{rgb}{0.88,1,1}
\newcolumntype{a}{>{\columncolor{Gray}}c}
\newcolumntype{b}{>{\columncolor{white}}c}

\def\asconv{\smash{\mathop{\longrightarrow}\limits^{as}}}   
   
\def\dconv{\smash{\mathop{\longrightarrow}\limits^d}}     

\def\max{\text{max}}
\def\min{\text{min}}
\def\min{\text{min}}
\def\var{\text{var}}

\def\diag\text{diag}
\def\thetab{\theta_{b}}

\def\thetabf{\theta_{b+1}}
\def\gd{\textsc{gd}}
\def\nr{\textsc{nr}}
\def\dmk{\textsc{dmk}}
\def\ks{\textsc{ks}}
\def\rnr{r\textsc{nr}}

\def\rgd{r\textsc{gd}}
\def\sgd{s\textsc{gd}}
\def\snr{\textsc{snr}}

\def\FL{\textsc{re}}
\def\MD{\textsc{md}}
\def\SMD{\textsc{smd}}
\def\bar{\overline}
\def\gnhat{G_n(\hat\theta_n)}
\def\pk{P_k}
\def\pb{P_b}

\def\hb{H_b}
\def\gk{G_k}
\def\gb{G_b}
\def\scriptgb{\mathbb G_b}
 
 \newtheorem{theorem}{Theorem}
 
 \newtheorem{proposition}{Proposition}
 \newtheorem{lemma}{Lemma}
 
 \newtheorem{assumption}{Assumption}

 \newcommand{\neutralize}[1]{\expandafter\let\csname c@#1\endcsname\count@}

 \newenvironment{assumptionbis}[1]
 {\renewcommand{\theassumption}{\ref*{#1}.iii$^\prime$}%
  \begin{assumption}}
 {\end{assumption}}

\begin{document}
\title{{\bf Inference by Stochastic Optimization:\\ A Free-Lunch Bootstrap}}
\author{Jean-Jacques Forneron\thanks{Department of Economics, Boston University, 270 Bay State Rd, MA 02215 Email: jjmf@bu.edu}  \and Serena Ng\thanks{Department of Economics, Columbia University and NBER, 420 W. 118 St. MC 3308, New York, NY 10027 Email: serena.ng@columbia.edu
\newline We would like to thank Jessie Li for helpful comments and suggestions as well as the participants of the Microeconomics Seminar at UC Santa-Cruz, the Optimization-Conscious Econometrics Conference,  and the econometric workshop held at BU/BC, Columbia University, University of Wisconsin, and New York University. We would also like to thank Robert Moffitt and Sisi Zhang for their help in replicating some of the empirical results as well as Wentian Qian for research assistance. Financial Support from the National Science Foundation (SES 1558623) is gratefully acknowledged.}
}
\date{September 2020}
\maketitle
\begin{abstract} 
Assessing sampling uncertainty in extremum estimation can be challenging when the asymptotic variance is not analytically tractable. Bootstrap inference offers a  feasible solution but  can be computationally costly especially when the model is complex. This paper uses iterates of a specially designed stochastic optimization algorithm as draws from which both point estimates and bootstrap standard errors can be computed \textit{in a single run}. The draws are generated by the gradient and Hessian computed from batches of data that are resampled at each iteration. We show that these draws yield consistent estimates and asymptotically valid frequentist inference for a large class of regular problems. The algorithm provides accurate standard errors in simulation examples and empirical applications at low computational costs. The draws from  the algorithm also provide a convenient way to detect  data irregularities. 

\end{abstract}


\setcounter{page}{0}
\thispagestyle{empty}
\baselineskip=18.pt
\newpage
\section{Introduction}
Many questions of economic interest can be expressed as non-linear functions of unknown parameters $\theta$ that need to be estimated from a sample of data of size $n$. The typical econometric routine is to first obtain a consistent estimate $\hat\theta_n$ of the true value $\theta^0$ by minimizing an objective function $Q_n(\theta)$, after which its sampling uncertainty is assessed. Though gradient-free optimizers provide point estimates, its asymptotic variance is often analytically intractable. One remedy is to use bootstrap standard errors, but this requires solving the minimization problem each time the data is resampled, and for complex models, this is no simple task. There is a long-standing interest in finding `short-cuts' that can relieve the computation burden without sacrificing too much accuracy. Examples include \citet{Davidson1999}, \citet{Andrews2002}, \citet{Kline2012}, \citet{Armstrong2014} and more recently \citet{Honore2017}. These methods provide standard errors by taking a converged estimate $\hat\theta_n$ as given. As such, estimation always precedes inference.

This paper proposes a resampling scheme that will deliver both the point estimates of $\theta$ and its standard errors  within the same optimization framework. Since the standard errors are obtained as a by-product of point estimation, we refer to the procedure as a `free-lunch bootstrap'.\footnote{In optimization, the no-free lunch theorem of \citet{wolpert-macready:97} states that, when averaged over all problems, the computation cost of finding a solution is the same across methods. We use the term to refer to the ability to compute the quantities for inference when the estimator is constructed.} The free-lunch is made possible by a specially designed stochastic optimization algorithm that resamples batches of data of size $m\le n$. Given an initial guess $\theta_0$, one updates $\thetab$ for $b\ge 0$ to $\theta_{b+1}$ using the gradient, the inverse Hessian as  conditioning matrix,  and a suitably chosen learning rate. We first show that the average over $B$ draws of $\thetab$ is equivalent to the mode $\hat\theta_n$ obtained by classical optimization up to order $\frac{1}{m}$. We then show that  the distribution of $\sqrt{m}(\theta_b-\hat\theta_n)$ conditional on the original sample of data is first-order equivalent to that of $\sqrt{n}(\hat\theta_n-\theta^0)$ upon rescaling, making it a bootstrap distribution. Because the conditioning matrix is the inverse Hessian, the procedure is a resampled Newton-Raphson (\rnr) algorithm.  For other conditioning matrices, the draws from resampling still produce a consistent estimate but cannot be used for inference.

The main appeal of the proposed methodology is its simplicity. If the optimization problem can be solved by our stochastic optimizer, inference can be made immediately without further computations. Natural applications include two-step estimation when the parameters in the two steps are functionally dependent in a complicated way, as well as minimum distance estimation that compares the empirical moments with the model moments expressed as a function of the parameters. When such a mapping cannot be expressed in closed-form, simulation estimation makes progress by using Monte-Carlo methods to approximate the binding function, but computing standard errors of the simulation-based estimates remains a daunting task. Our algorithm provides an automated solution to compute standard errors and removes simulation noise, resulting in more accurate estimates. The algorithm also provides a convenient way to compute clustered standard errors and model diagnostics. 

As compared to other stochastic optimization algorithms, we use a learning rate that is fixed rather than vanishing, and though a small $m$ is desirable from a pure computation perspective, valid inference necessitates that $m$ cannot be too small. As compared to conventional bootstrap methods, the simultaneous nature of estimation and inference means that a preliminary $\hat\theta_n$ is not needed for resampling. Though $\thetab$ is a Markov chain, no prior distribution is required, nor are Bayesian computation tools employed. In simulated examples and applications, our bootstrap standard errors match up well with the asymptotic and bootstrap analogs, but at significantly lower computational costs.

The plan of the paper is as follows. Section \ref{sec:classical} begins with a review of classical and stochastic optimization. The proposed free-lunch algorithm is presented in Section \ref{sec:rnr} and its relation to other resampling procedures is explained. The properties of the draws from the algorithm are derived in Section \ref{sec:theory}. Simulated and empirical examples are presented  in Section \ref{sec:examples}. Section \ref{sec:smd} extends the main results to simulation-based estimation. Appendix \ref{apx:proofs} provides  derivations of the main results.  An on-line  supplement\footnote{The file  is available  for download at \underline{\textcolor{blue}{\url{www.columbia.edu/~sn2294/papers/freelunch-supp.pdf}}}.} provides the  \textsc{r} code to implement one of the  applications considered,  additional  results  with details for replications,  as well as  an analytical example for Section \ref{sec:smd}.

\section{Review of the Related Literature} \label{sec:classical}
Consider minimization of the objective function $Q_n(\theta)$ with respect to $\theta$ whose true value is $\theta^0$. The sample gradient and Hessian of $Q_n(\theta)$ are defined respectively by
 \begin{eqnarray*}
G_n(\theta)&=&\nabla Q_n(\theta;x)=\frac{1}{n}\sum_{i=1}^n \nabla Q_n(\theta;x_i) \\
H_n(\theta)&=&\nabla^2 Q_n(\theta;x)=\frac{1}{n}\sum_{i=1}^n \nabla^2 Q_n(\theta;x_i).
\end{eqnarray*} 
 The necessary conditions for a local minimum are $\|G_n(\hat\theta_n)\|=0$ and $H_n(\hat\theta_n)$ positive semi-definite. The sufficient conditions are $\|G_n(\hat\theta_n)\|=0$ and $H_n(\hat\theta_n)$ positive definite. To find the optimal solution, a generic rule for updating from the current estimate $\theta_k$ is 
\[ \theta_{k+1}=\theta_k -\gamma_k Z_n(\theta_k)\]
where $\gamma_k$ is the step size and $Z_n=\frac{\partial \theta_{k+1}}{\partial \gamma_k}$
 is the direction of change. 

 Gradient based methods specify $Z_n(\theta_k)= P_n(\theta_k) G_n(\theta_k)$ where $P_n(\theta_n)$ is a conditioning matrix. The updating rule then becomes 
  \begin{eqnarray}
   \theta_{k+1}         &\equiv &\theta_k-\gamma_k  P_n(\theta_k) G_n(\theta_k).
\label{eq:gradient-rule}
 \end{eqnarray}

The method of gradient descent (\textsc{gd}) (also known as steepest descent) sets $ P_n=I_d$. Since \textsc{gd} does not involve the Hessian, it is a first order method and is less costly. Convergence of $\hat\theta_k$ to the minimizer $\hat\theta_n$ is linear under certain conditions,\footnote{In statistical computing, the convergence of $\theta_k$ to $\hat\theta_n$ is said to be linear if $\|\theta_{k+1}-\hat\theta_n\|/\|\theta_k-\hat\theta_n\|^q<r$ for some $r\in (0,1)$ if $q=1$ and quadratic if $q=2$. Convergence is superlinear if $\lim_{k\rightarrow\infty} \|\theta_{k+1}-\hat\theta_n\|/ \|\theta_k-\hat\theta_n\|=0$. See \citet{boyd-vandenberghe:04} Section 9.3.1 for linear convergence of gradient methods, and \citet[Theorem 3.5]{nocedal-wright:06} for quadratic convergence of Newton's method when $\gamma=1$ or $\gamma_k \to 1$ at an appropriate rate.   `Damped Newton' updating with  $\gamma_k \in (0,1)$ has a linear convergence rate, see \citet{boyd-vandenberghe:04} Section 9.5.3 and \citet{nesterov2018} Section 1.2.4.} but the rate depends on $I_d-\gamma H_n(\theta_k)$  being in the restricted region of $(-1,1)$, and can be slow when the ratio of the maximum to the minimum eigenvalue of $H_n$ is large.
The Newton-Raphson algorithm puts $P_n=H_n(\theta_k)^{-1}$. It is    a second-order method since it involves the Hessian matrix.
 When $\gamma_k=1$, the algorithm converges quadratically if $Q_n$ satisfies certain conditions. A drawback of Newton's algorithm is that it requires computation of the inverse of the Hessian. When strong convexity fails, the Hessian could be non-positive definite for $\theta$ away from the minimum. In such cases, it is not uncommon to replace the Hessian  by $H_n(\theta_k)+c\cdot I_d$ for some $c>0$, or specify $P_n=(H_n(\theta_k)^\prime H_n(\theta_k))^{1/2}$ to restore positive definiteness around saddle-points, see \citet[Chapter 3.4]{nocedal-wright:06}. Quasi-Newton methods bypass direct computation of the Hessian or its inverse, but analytical convergence results are more difficult to obtain. We  focus our theoretical analysis on  gradient descent and Newton-Raphson based algorithms but will consider quasi-Newton methods in some simulations.

\subsection{Stochastic Optimization}

 Stochastic optimization finds the optima in noisy observations using carefully designed recursive algorithms. The idea can be traced to the theory of stochastic approximation when the goal is to minimize some function $Q(\theta)$ with gradient $G(\theta)$,  which is equivalent to the root-finding problem  $G(\theta)=0$ whose the true value is $\theta^0$. A classical optimizer would perform
$ \theta_{k+1}=\theta_{k}-\gamma_{k} G(\theta_{k})$. \citet{Robbins1951} considers the situation when we only observe $G(\theta_{k})+e_k$ with $\mathbb E(e_k)=0$ and suggests to update according to
\[ \theta_{k+1}=\theta_{k}- \gamma_{k} (G(\theta_{k})+e_k).\]
 \citet{Robbins1951} proved that $\theta_k\asconv \theta^0$ for $G$ non-decreasing with step size sequence $\gamma_k \geq 0$ satisfying
\begin{equation}
\label{eq:munro-robbins-conditions}
 \text{(i)}\quad \sum_{k=1}^\infty \gamma_k= +\infty, \quad\quad   \text{(ii)} \quad\sum_{k=1}^\infty \gamma_k^2<+\infty.
\end{equation}
The first condition ensures that all possible solutions will be reached with high probability regardless of the starting value, while the second ensures convergence to the true value. Building on the Robbins-Monro algorithm, the Kiefer-Wolfowitz algorithm uses a finite difference approximation 
$G(\theta_k)\approx G_n(\theta_k)=\frac{1}{2\epsilon_k}\bigg[ Q_n(\theta_k+\epsilon_k)-Q_n(\theta_k-\epsilon_k)\bigg]$. This is often recognized as the first implementation of stochastic gradient descent. \citet{Kiefer1952Stochastic}
 proves convergence of $\theta_k$ to the maxiumum likelihood estimate $\hat\theta_n$ assuming that the likelihood $Q_n$ is convex, $\epsilon_k$ goes to zero, and that the two conditions stated in (\ref{eq:munro-robbins-conditions}) hold.


Modern  stochastic gradient  descent updates according to
\begin{eqnarray*}
       \theta_{k+1}&=&\theta_{k}- \gamma_k G_m(\theta_{k})
\end{eqnarray*}
where $G_m(\theta_k)=\frac{1}{m}\sum_{i=1}^m G(\theta_k;x_i)$ is an estimate of $G(\theta)$.
It can be seen as Monte-Carlo based since the $m$ observations used to compute $G_m(\theta_k)$ are usually chosen from $\{1,\dots,n\}$  randomly.  Though $m=1$ is computationally inexpensive and is a popular choice, a small $\gamma_k$ is often needed to compensate for the higher variation. A common rule is to choose $\gamma_k = \gamma k^{-\delta}$, where $\delta \in (1/2,1]$ and $\gamma>0$ are the choice parameters. Depending on $\delta$, convergence as measured by $\mathbb{E}(\|\theta_k-\theta^0\|^2)$ can occur at a $1/k$ rate or slower. To reduce sensitivity to the tuning parameters, \citet{ruppert1988efficient} and \citet{polyak1992acceleration} propose to accelerate convergence using what is now known as Polyak-Ruppert averaging: $\overline{\theta}_k = \frac{1}{k}\sum_{i=1}^k \theta_i$. Importantly, $\overline{\theta}_k$ converges at the fastest $1/k$ rate for all choices of $\delta \in (1/2,1]$. \citet{Bach2011} 
shows that the improvements hold even for a finite number of iterations $k$.  We will return to  Polyak-Ruppert averaging below.

Stochastic optimization presents an interesting alternative to classical optimization as it approximates the gradient on  {\it minibatches} of the original data. This is particularly helpful in large scale learning  problems  such Lasso, support-vector machines and K-means clustering when non-linear optimization   can be challenging. Improvements to \textsc{sgd} with $P_m=I_d$ include  momentum \citep{polyak:64} and accelerated gradient \citep{nesterov:83} methods.  
 Besides its computational appeal, stochastic optimization can improve upon its classical counterpart in non-convex settings.\footnote{See \citet[Chapter 8]{Goodfellow-et-al-2016} for an overview of \textsc{sgd}. \citet{ge2015escaping} shows that noisy gradient descent can escape all saddle points in polynomial time under a strict saddle property whereas classical gradient methods converge at saddle points where the gradient is zero. \citet{jin2017escape} shows that the dimension of $\theta$ has a negligible effect on the number of iterations needed to escape saddle points, making it an effective solution even in large optimization problems.}

A variation of \sgd, known as Stochastic gradient Langevin dynamics (\textsc{sgld}) incorporates Langevin dynamics into a Bayesian sampler.  As will be discussed further below, the update is based on the gradient of the posterior distribution.
Of note now is that   \textsc{sgld}  has two types of noises: an injected noise,  and the stochastic gradient noise based on $m \ll n$ observations. They play different roles in the algorithm. In the early phase of \textsc{sgld}, the stochastic gradient dominates and the algorithm performs optimization. In the second phase, the injected noise  dominates and the algorithm behaves like a posterior sampler. The algorithm seamlessly switches from one phase to another for an appropriate choice of the learning rate. 

Unlike classical optimization, stochastic Newton-Raphson with the inverse Hessian  $H_m(\theta)$ as conditioning matrix is not popular because the Hessian is often noisy and  near singular for $m$ small, rendering the algorithm unstable. We will show that using a variation of stochastic Newton-Raphson with larger batches of data can produce draws that not only provide an accurate estimate of $\theta^0$ but also yields frequentist assessment of sampling uncertainty.  It thus integrates numerical optimization with statistical inference. In contrast, other conditioning matrices will yield consistent estimates but would not provide valid inference in our setup.

\section{Extremum Estimation and Inference by Resampling} \label{sec:rnr}
Consider extremum estimation of parameters $\theta\in\Theta \subset \mathbb{R}^d$ from data  $x=(x_1,\ldots, x_n)$. Let $\theta^0$ be the minimizer of a twice differentiable population objective function $Q(\theta)$ whose sample analog is 
$ Q_n(\theta)\equiv Q_n(\theta;x)$.  The sample extremum estimator is
\[ \hat\theta_n=\text{argmin}_{\theta \in \Theta} Q_n(\theta).\]
  For likelihood estimation, $Q_n(\theta)=-\sum_{i=1}^n \ell_i(\theta)$
 where  $\ell_i$ is the likelihood of $\theta$ at observation $x_i$. For least squares estimation, $Q_n(\theta)$ is the sum of squared residuals $\sum_{i=1}^n e_i^2(\theta)$. For GMM estimate with positive weighting matrix $W_n$, $Q_n(\theta)=\bar{g}_n(\theta)^\prime W_n \bar{g}_n(\theta)$ where $\mathbb E [g_i(\theta^0)]=0$.   Under regularity conditions stated in Theorem 2.1 of  \citet{newey-mcfadden-handbook}   $\hat\theta_n$  is consistent for $\theta^0$. If, in addition, the assumptions in Theorem 3.1 of \citet{newey-mcfadden-handbook} hold, then $\hat\theta_n$ is also $\sqrt{n}$-asymptotically normal:
 \[\sqrt{n}(\mathbb{V}^{0})^{-1/2}(\hat\theta_n-\theta^0)\dconv  N(0,I_d) \]
where $\mathbb V^0=[H(\theta^0)]^{-1} \text{var}(\sqrt{n}G_n(\theta^0)) [H(\theta^0)]^{-1}$. Finite sample inference is typically based on an estimate  of $\mathbb V^0$ which    can be analytically intractable or costly  to compute on the full sample.  It is not uncommon to resort to  bootstrap inference. We consider  the $m$ out of $n$ bootstrap with $m \to \infty, m/n \to c\in [0,1]$ and samples $(x_1^{(b)},\dots,x_{m}^{(b)})$ with replacement from the data $(x_1,\dots,x_n)$ for $b=1,\dots,B$ and solve $B$ minimization problems:
\[ \hat \theta_m^{(b)}  =\text{argmin}_{\theta \in \Theta} Q_m^{(b)}(\theta), \]
where the resampled objective $Q_m^{(b)}(\theta)=Q_m^{(b)}(\theta,x^{(b)})$ is computed over the sample $x^{(b)}$. 

Let $\mathbb{E}^\star$ and $\text{var}^\star$ denote the bootstrap expectation and variance which are taken conditional on the sample data $(x_1,\dots,x_n)$, and $\overset{d^\star}{\to}$ denotes the convergence in distribution conditional on the data. Since we only consider correctly specified regular estimators,   the desired convergence is to a Gaussian limit. The $m$ out of $n$ bootstrap can allow for different sampling schemes.   Assuming that the  resampling scheme is chosen to reflect the  dependence structure of the data, it holds in a variety of settings that\footnote{For two-way clustering, a recommended procedure is to resample over one cluster dimension and reweigh along the other \citep{roodman2019fast}. See also \citet{cameron2011robust} on multiway clustering. For time-series data, block resampling is needed to preserve the dependence structure. For correctly specified GMM models the bootstrap described above is valid \citep{hahn1996}. 
} 
\[ \sqrt{m} (\mathbb{V}_m)^{-1/2} \left( \hat\theta_m^{(b)} - \hat\theta_n \right) \overset{d^\star}{\to} \mathcal{N} \left( 0, I_d \right),\]
 where $\mathbb V_m=[H_n(\hat\theta_n)]^{-1} \text{var}^\star(\sqrt{m}G_m^{(b)}(\hat\theta_n)) [H_n(\hat\theta_n)]^{-1}$  depends on the inverse Hessian as well as the variance of the resampled score. The result implies that the resampled distribution of $\hat\theta_m^{(b)}$ can be used to approximate the sampling distribution of $\hat\theta_n$.   

An $m$ out of $n$ bootstrap with $m<n$ uses smaller samples  and performs as well as  a $n$ out of $n$ bootstrap in simulations while  requiring similar or weaker conditions, \citep{bickel2012}. Nonetheless, it still makes multiple calls to a classical optimizer. 
As will be seen below, our proposed algorithm only requires one call to the optimizer.



We propose the following two algorithms and use `$b$' to index the iterates of the algorithm.  In this notation, $G_m^{(b+1)}(\thetab)$ is the gradient computed on the $(b+1)$-th batch of resampled data of size $m$ and evaluated at $\thetab$, the parameter value in the previous draw, and $H_m^{(b+1)}(\thetab)$ is similarly defined.

\begin{algorithm}[H] 
\caption{Estimation by Resampling} \label{algo:oe} 
      \begin{algorithmic}
        \State \textbf{Input:} (a) an initial guess $\theta_{0}$, (b) a bootstrap sample size $B$ and a burn-in period \textsc{burn}, (c) a batch size $m\leq n$,   (d) a  fixed learning rate $\gamma \in (0,1]$, (e) a conditioning matrix $P_b$
        \State \textbf{Burn-in and Resample:} 
        \For{$b=1,\dots,\textsc{burn}+B$} 
            \State Resample the $(b+1)$-th batch of data of size $m$
            \State Update  $\pb $ and $G_b= G_m^{(b+1)}(\thetab)$.
            \State Update $\theta_{b+1} = \thetab - \gamma \pb \gb,$
      \EndFor
 \State \textbf{Discard} the first \textsc{burn} draws, re-index $\theta_{\text{burn}+b}$ to $\thetab$ for $b=1,\ldots B$.
\State Let  $\overline{\theta}_{\FL}=\frac{1}{B} \sum_{b = 1}^B \thetab$.
      \end{algorithmic}
\end{algorithm}

\bigskip

\begin{algorithm}[H] 
 \caption{The Free-Lunch Bootstrap} \label{algo:free_lunch}
      \begin{algorithmic}
        \State Implement Algorithm \ref{algo:oe} with $P_b=H_b^{-1}$.

\State Let  $\overline{\theta}_{\rnr}=\frac{1}{B} \sum_{b = 1}^B \thetab$ and  define 
$ \widehat{\var}(\theta_b) =\frac{1}{B} \sum_{b = 1}^B (\thetab-\overline{\theta}_{\rnr})(\thetab-\overline{\theta}_{\rnr})^\prime.$
      \State \textbf{Output:}    
$\overline{\theta}_{\rnr}$ and
$V_{\rnr} = \frac{m}{\phi(\gamma)}\widehat\var(\theta_b)$, where $\phi(\gamma)=\frac{\gamma^2}{1-(1-\gamma)^2}$.\\
      \end{algorithmic}
\end{algorithm}


Algorithm \ref{algo:oe} produces an estimate $\bar\theta_{\FL}$ by resampling, hence the acronym \FL. It  works for any conditioning matrix $P_b$ satisfying assumptions to be made precise in Theorem \ref{th:average_estimation}. 
Algorithm \ref{algo:free_lunch} produces draws using the inverse Hessian as $P_b$ as in  Newton-Raphson, hence the acronym  \rnr.  The free-lunch aspect  relates to the fact that we get both an estimate $\overline{\theta}_{\rnr}$ and its standard error in one run.
The bootstrap aspect comes from the fact that under the assumptions of Theorem \ref{th:free_lunch}, 
 $\sqrt{m}(\theta_b-\hat\theta_n)$ has the same asymptotic distribution as $\sqrt{n}(\hat\theta_n-\theta^0)$ after  an adjustment of $\frac{m}{\phi(\gamma)}$.  The quantity $V_{\rnr}$  is an estimate of the sandwich variance that is computationally costly for classical estimation.  A Wald test for $H_0:\theta=\theta^\dag$ can be constructed as
\[ \text{wald}= n(\overline\theta_{\rnr}-\theta^\dag)^\prime V_{\rnr}^ {-1} (\overline\theta_{\rnr}-\theta^\dag)\]
which has an asymptotic Chi-squared distribution under the null hypothesis. A 95\% level bootstrap confidence interval can also be constructed after adjusting for $\frac{m}{n}$ and $\phi(\gamma)$ by taking the (0.025,0.975) quantiles of $\big\{ \overline{\theta}_{\rnr}+\sqrt{\frac{m}{n\phi(\gamma)}}(\thetab - \overline{\theta}_{\rnr})\big\}_{b \geq 1}$.



Algorithms \ref{algo:oe} and \ref{algo:free_lunch} have three features that distinguish them from existing gradient-based stochastic optimizers. First, $\gamma\in (0,1]$ does not change with $b$. 
Fixing $\gamma$ rather than letting $\gamma_b\rightarrow 0$ potentially permits faster convergence. Second, we sample $m$ out of $n$ observations with $m/n\rightarrow c\in [0,1]$ and $\sqrt{n}/m \to 0$. This precludes the popular choice in stochastic optimization of $m=1$, but admits $m=n$. We thus accept a higher computation cost to accommodate inference. Third, compared to \sgd\,  Algorithm \ref{algo:free_lunch} uses the inverse Hessian as conditioning matrix.

\subsection{The Linear Regression Model}
This subsection uses the linear regression model to gain intuition of the Free-Lunch bootstrap. 
The  model is  $y_i=x_i^\prime\theta+e_i$. Let $\hat e_n = y_n-X_n \hat\theta_n$ be the $n\times 1$ vector of least squares residuals evaluated at the solution $\hat\theta_n$. $X_n$ denote the $n\times K$ matrix of regressors. The linear model is of interest because the objective function is quadratic and the quantities required for updating are analytically tractable. The gradient and  Hessian of the full sample objective function
$ Q_n(\theta) =(y_n-X_n\theta)^\prime(y_n-X_n\theta)/(2n)$ are
$ G_n(\theta) = -X_n^\prime (y_n-X_n\theta)/n$ and $H_n(\theta) = X_n^\prime X_n/n$. The updates for this linear model  evolve as  
\begin{eqnarray*}
\theta_{k+1}&=&  \theta_{k}+\gamma P_{k} X_n^\prime(y_n-X_n\theta_k)/n\\
&=& \theta_{k}+\gamma P_{k} X_n^\prime\bigg(X_n(\hat\theta_n-\theta_{k})+\hat e_n\bigg)/n.
\end{eqnarray*}
 Convergence of $\theta_k$  for a given conditioning matrix $P_k$ can be studied by subtracting  $\hat\theta_n$ from both sides of the updating equation and re-arranging terms (see Appendix \ref{apx:OLS} for details). Table \ref{tbl:penpencil} summarizes convergence of  \gd, \nr, \sgd, \rgd\, and  \rnr. \snr\, is not considered because $X_m^\prime X_m/m$ is singular for $m=1$ so $\theta_b$ is not well defined. The left panel of the table gives the updating rule in closed form and the right panel expresses the deviation of the draws from $\hat\theta_n$ as the sum of a deterministic and a stochastic component.  

\begin{table}[ht]
\caption{OLS: updating rules and convergence}
\label{tbl:penpencil}
  \centering
    \begin{tabular}{l|c|c|rcl} \hline \hline
      \multirow{2}{*}{Method} & Conditioning & Update:   & \multicolumn{3}{c}{Convergence: $\theta_{k+1}-\hat\theta_n$=}\\  
&Matrix $\pk$ & $\theta_{k+1}-\theta_k=$ &  deterministic &+ &  stochastic\\ \hline
      \textsc{gd} & $I_d$ &$ -\gamma_k G_k $ & $(I_d-\gamma H_n)(\theta_k-\hat\theta_n)$ && \\     
      s\textsc{gd} & $I_d$ & $-\gamma_b G_b$ & $(I_d-\gamma_b H_b )(\thetab-\hat\theta_n)$ &$-$& $\gamma_b G_b(\hat\theta_n)$\\
      r\textsc{gd} & $I_d$ & $-\gamma G_b$ & $(I_d-\gamma H_b )(\thetab-\hat\theta_n)$ &$+$& $\gamma H_b(\hat\theta_m^{(b+1)}-\hat\theta_n)$\\  \hline
      \textsc{nr} & $H_k^{-1}$ &  $ - \gamma H_k^{-1} G_k $ & $(1-\gamma) (\theta_k-\hat\theta_n)$ & & \\    
 \rnr & $H_b^{-1}$ &
           $  -\gamma  H_b^{-1}   G_b$ & $ (1-\gamma) (\thetab-\hat\theta_n)$ & $+$ & $\gamma (\hat\theta_m^{(b+1)}-\hat\theta_n)$\\
 \hline \hline        
 \end{tabular}\\
\textit{Note: $G_k=G_n^{(k+1)}(\theta_k)$, $G_{b}=G_m^{(b+1)}(\thetab)$, $H_k=H_n^{(k+1)}(\theta_k)$, $H_b=H_m^{(b+1)}(\thetab)$. }
\end{table}

As seen from Table \ref{tbl:penpencil}, \textsc{gd} updates do not depend on the Hessian but convergence does, while for \nr\, the opposite is true. Convergence of \nr\;  can be achieved after one iteration if $\gamma=1$.  In \sgd, \rgd\, and \rnr, batch resampling adds a stochastic component to the updates and convergence is no longer deterministic. 
The deviations for \sgd\, and \rgd\, $\thetabf-\hat\theta_n$ follow a VAR(1) process with varying and fixed coefficient matrices $I_d-\gamma_b H_b$ and $I_d-\gamma H_b$, respectively. 
In contrast, the \rnr\ draws have an AR(1) representation with a fixed coefficient $(1-\gamma)$ that is dimension-free and independent of the Hessian. Note that \rgd\, and \rnr\, keep $\gamma$ fixed and rely on averaging over $b$ for convergence. 

Our main result pertains to \rnr\; so it is useful to  have a deeper understanding of how it works.
Unlike  stochastic optimizers which require $\gamma_b$ vanishing, the learning rate $\gamma$ used to generate the  \rnr\, draws  is constant. The   draws evolve according to
\begin{equation}
\label{eq:rnr-ols} \thetabf-\hat\theta_n= (1-\gamma)(\thetab-\hat\theta_n)+\gamma (\hat\theta^{(b+1)}_m-\hat\theta_n)
\end{equation}
where  $\hat\theta_m^{(b+1)}=(X_m^{(b+1)\prime} X_m^{(b+1)})^{-1}X_m^{(b+1)\prime} y_m^{(b+1)}$ is obtained by classical optimization using the $(b+1)$-th bootstrap sample $(y_i^{(b+1)},x_{i}^{(b+1)})_{i=1,\dots,m}$.  Being a bootstrap estimate, it holds  under regularity conditions that the distribution of $\sqrt{m}(\hat\theta_m^{(b+1)}-\hat\theta_n)$ conditional on the data approximates the sampling distribution of $\sqrt{n}(\hat\theta_n-\theta^0) \dconv N(0,\mathbb V^0)$.

Clearly when $\gamma=1$, (\ref{eq:rnr-ols}) implies   $\thetab = \hat\theta_m^{(b)}$, meaning that each \rnr\; draw  equals  the  bootstrap estimate $\hat\theta_m^{(b)}$. We want to show that the draws are still bootstrap estimates  when $\gamma\in(0,1)$. For such $\gamma$,  $\thetabf-\hat\theta_n$ is an AR(1) process where for each $b$, the innovations $ \gamma(\hat\theta_m^{(b+1)}-\hat\theta_n)$ are iid conditional on the original sample. 
Iterating the AR(1) formula backwards to the initial value $\theta_0$, we can decompose the draws $\thetabf$ into two terms:
\begin{align}
      \thetabf - \hat\theta_n=  \underbrace{  \vphantom{ \gamma \sum_{j=0}^b (1-\gamma)^j (\hat\theta_m^{(b+1-j)}-\hat\theta_n) }  (1-\gamma)^{b+1}(\theta_0 - \hat\theta_n)}_{\text{initialization bias}} + \underbrace{\gamma \sum_{j=0}^b (1-\gamma)^j (\hat\theta_m^{(b+1-j)}-\hat\theta_n)}_{\text{resampling noise}}, \label{eq:AR1_OLS}
\end{align}
 where $\{\hat\theta_m^{(b+1-j)}\}_{j \geq 0}$ are the bootstrap estimates in the previous iterations. The constant learning rate is crucial in achieving this representation.

To show that our estimator $\overline\theta_{\rnr}$ is $\sqrt{n}$-consistent for $\hat\theta_n$, i.e. $\bar \theta_{\rnr}=\hat\theta_n+o_{p^\star}(\frac{1}{\sqrt{n}})$, we need to evaluate the average of the two terms in (\ref{eq:AR1_OLS}) over $b$.  The initialization bias  in (\ref{eq:AR1_OLS})   is  due to taking an arbitrary starting value $\theta_0$ and is identical to the optimization error in classical Newton-Raphson.  For $\gamma\in(0,1]$, $\frac{1}{B}\sum_{b=1}^B (1-\gamma)^{b+1}=O(\frac{1}{B})$ because $\{ (1-\gamma)^{b} \}_{b \ge 1}$ is a summable geometric series. Another bias term of order $O(\frac{1}{m})$ arises when $\mathbb{E}^\star(\hat\theta_m^{(b)}-\hat\theta_n)=O(\frac{1}{m})$.  Since $\hat\theta_n$ is fixed as $b$ varies,  we now have
$\mathbb E^\star (\overline{\theta}_{\rnr})=\hat\theta_n+O(\frac{1}{B}) +O(\frac{1}{m}).$
Thus $\mathbb E^\star (\overline{\theta}_{\rnr})=\hat\theta_n +o(\frac{1}{\sqrt{n}})$  as required, assuming $\frac{\sqrt{n}}{\min(m,B)}\rightarrow 0$. Turning to the variance, first note that by virtue of bootstrapping, $\{\hat\theta_m^{(b+1-j)}-\hat\theta_n\}_{j \geq 0}$ constitutes a sequence of conditionally iid errors each with variance that is $O(\frac{1}{m})$. Since   $\{ (1-\gamma)^{b} \}_{b \geq 1}$ is summable, the   variance in $\overline{\theta}_{\rnr}$ due to resampling is $O(\frac{1}{mB})$. This becomes $o(\frac{1}{n})$ when  $\frac{n}{mB}\rightarrow 0$, a sufficient condition being $\frac{\sqrt{n}}{\min(m,B)} \to 0$, which is also required for the bias to be negligible. We have thus shown  that $\overline{\theta}_{\rnr} = \hat\theta_n + o_{p^\star}(\frac{1}{\sqrt{n}})$ for $\frac{\sqrt{n}}{\min(m,B)} \to 0$, which is a simplified version of Theorem \ref{th:average_estimation} below.  Though the result has the flavor of   Polyak-Ruppert averaging in stochastic optimization,  $\gamma$ is fixed here and $m$ increases with $n$.

To show bootstrap validity of \rnr, we need to establish that, conditional on the sample of data, the distribution of $\sqrt{m}(\theta_{b+1}-\hat\theta_n)$  is asympotically equal, up to a constant scaling factor, to that of $\sqrt{m}(\hat\theta_m^{(b+1)}-\hat\theta_n)$. This  requires  that the initialization bias in each $\thetab$   is $o(\frac{1}{\sqrt{m}})$, which holds when $\frac{\log(m)}{b} \to 0$.  From (\ref{eq:AR1_OLS}), $\sqrt{m}(\theta_{b+1}-\hat\theta_n)$  has variance $\gamma^2 \mathbb{V}_{m}$ conditional on $\thetab$ and unconditional variance \[\var\bigg(\sqrt{m}(\theta_{b+1}-\hat\theta_n)\bigg)=\frac{\gamma^2 + O([1-\gamma]^{b+2})}{1-[1-\gamma]^2}\mathbb{V}_{m}\approx \phi(\gamma)\mathbb V_m\]
where $\phi(\gamma) = \frac{\gamma^2}{1-[1-\gamma]^2}$, and  $\mathbb V_m=\var(\sqrt{m}(\hat\theta^{(b+1)}_m-\hat\theta_n))$ is the bootstrap estimate of the sandwich variance $\mathbb{V}^0$ defined above. This
 establishes  that the variance of $\thetab$ is proportional to that of the bootstrap estimate. As shown in \citet{gonccalveswhite2005}, $\mathbb V_m$  is consistent for $\mathbb V^0$ under certain moment conditions. This implies that, up to the scaling factor $\phi(\gamma)$, the co-variance of $\theta_b$ is also consistent for $\mathbb V^0$.  Combined with  asympotic normality of each $\sqrt{m}(\hat\theta_m^{(b+1)}-\hat\theta_n)$ for each $b$ and  additional conditions to be made precise in Theorem \ref{th:free_lunch}, we have
 \[  \bigg(\phi(\gamma)\mathbb{V}_{m}\bigg)^{-1/2}\sqrt{m}  \left( \thetabf - \hat \theta_n \right) \overset{d^\star}{\to} \mathcal{N} \left( 0, I_d\right). \]
 But asymptotic theory gives the distribution of  $\sqrt{n}(\hat\theta_n-\theta^0)$ with sample size $n$, not $m$. An adjustment for $\phi(\gamma)$ and $m$ is needed.  
Let  $\mathbb V_{\rnr}= \frac{m}{\phi(\gamma)}\text{var}^\star(\thetab) = \mathbb{V}_m + o(1)$. For appropriate choice of $m$ and $\gamma$,
$ \mathbb V_{\rnr}^{-1/2} \sqrt{n}(\overline\theta_{\rnr}-\theta^0) \overset{d^\star}{\to} \mathcal N(0,I_d) $ and
 Algorithm \ref{algo:free_lunch} proposes a plug-in estimate of $\mathbb V_{\rnr}$.

\section{Properties of the Draws $\thetab$} \label{sec:theory}

This section studies the properties of  draws $\thetab$ produced by Algorithms \ref{algo:oe} and \ref{algo:free_lunch} for non-linear models. The proofs are more involved for two reasons. First,
an  arbitrary $\gamma \in (0,1]$ may not lead to convergence even for classical optimizers. Second, whereas in quadratic problems the draws $\thetab$ have a tractable AR(1) representation, for non-quadratic objectives the draws $\thetab$ follow a non-linear process which is more difficult to study.  
Hence we need to first show, under strong convexity conditions, that there exist fixed values of $\gamma \in(0,1]$ such that optimization of $Q_n(\theta)$ has a globally convergent solution. We then show, using the idea of coupling, for  appropriate choices of $m$ and $B$ that $\thetabf$ can be made very close to a linear AR(1) sequence $\thetabf^\star$ that is constructed as if the objective were quadratic. This allow us to establish  consistency of $\bar\theta_{\FL}$ for $\hat \theta_n$  in Theorem \ref{th:average_estimation} for a large class of $\pb$, and a distribution result in Theorem \ref{th:free_lunch} that validates inference for a particular choice of  $\pb$.

\subsection{Convergence of $\theta_k$ to $\hat\theta_n$ from Classical Updating}
Econometric theory typically studies the conditions under which $\hat\theta_n$ is consistent for $\theta^0$, taking as given that a numerical optimizer exists to produce a convergent solution $\hat\theta_n$. From \citet{newey-mcfadden-handbook}, the regularity conditions for consistent estimation of $\theta$  are continuity of $Q(\theta)$ and uniform convergence of $Q_n(\theta)$ to $Q(\theta)$.  Asymptotic normality further requires smoothness of $Q_n(\theta)$,  $\theta^0$ being in the interior of the support, and non-singularity of  $H(\theta^0)$.  But classical Newton-type algorithms may only converge to a local minimum and a global convergent solution is guaranteed only when the objective function is strongly convex on the parameter space $\Theta$.  For gradient-based optimizers to deliver such a solution, the following provides the required conditions.

\begin{assumption} \label{ass:Qn}
      $Q_n$ is twice continuously differentiable on $\Theta$, a convex and compact subset of $\mathbb{R}^d$.   There exists a constant $C_1 < +\infty$ such that for all $\theta \in \Theta$:
      \begin{itemize}
            \item[i.] $0<\underline{\lambda}_H \leq \lambda_{\min}(H_n(\theta)) \leq \lambda_{\max}(H_n(\theta)) \leq \overline{\lambda}_H<+\infty$,
            \item[ii.] $\| H_n(\theta) - H_n(\hat \theta_n) \|_2 \leq  C_{1}  \|\theta - \hat \theta_n\|_2$,
            \item[iii.]  $0<\underline{\lambda}_P \leq \lambda_{\min} (\pk) \leq \lambda_{\max}(\pk)\leq \overline{\lambda}_P<+\infty.$
      \end{itemize} 
\end{assumption}
Condition i. implies strong convexity of $Q_n$ on $\Theta$.\footnote{See \citet{boyd-vandenberghe:04}, Chapter 9.1. A function $Q_n$ is strongly convex on $\Theta$ if for all $\theta\in\Theta$,  there exists some $\underline{\lambda}>0$ such that $\nabla^2 Q_n(\theta) \ge \underline{\lambda} I_d$. For bounded $\Theta$, there also exists $\overline{\lambda}$ such that  $\nabla^2 Q_n(\theta) \le \overline{\lambda} I_d$. Then $\overline{\lambda}/\underline{\lambda}$ is an upper bound on the condition number of $\nabla^2 Q_n(\theta)$.}  Condition ii. imposes Lipschitz continuity of the Hessian.
 Assumption \ref{ass:Qn} implies the following two inequalities which are known as the Polyak-\L{}ojasiewicz inequalities:
\begin{align} 
\label{eq:cvx1} \langle \theta - \hat \theta_n, \gnhat \rangle &= (\theta - \hat \theta_n)^\prime H_n(\tilde \theta_n) (\theta - \hat \theta_n) \geq \underline{\lambda}_{H}  \|\theta-\hat\theta_n\|_2^2, \\
\label{eq:cvx2}
      \|\gnhat\|_2^2 &=  (\theta - \hat \theta_n)^\prime H_n(\tilde \theta_n)^2 (\theta - \hat \theta_n) \leq \overline{\lambda}_H^2  \|\theta-\hat\theta_n\|^2_2,
\end{align}
where $\tilde{\theta}_n$ is an intermediate value between $\theta$ and $\hat{\theta}_n$. Inequality (\ref{eq:cvx1}), due to \citet{lojasiewicz1963} and \citet{polyak1963}, follows from the  positive definiteness of $H_n(\tilde \theta_n)$. Together, (\ref{eq:cvx1}) and (\ref{eq:cvx2}) ensure that $\hat \theta_n$ is a unique (or global) minimizer of $Q_n$. 

Assumption \ref{ass:Qn} also implies that there exists $\gamma$ such that  gradient based optimization is globally convergent. To see why, consider 
\begin{align*} \label{eq:contract}
      \|\theta_{k+1}-\hat\theta_n\|_2^2 &= \|\theta_{k}-\hat\theta_n - \gamma \pk \gk\|_2^2 \nonumber\\
      &= \|\theta_{k}-\hat\theta_n\|_2^2 - 2 \gamma  \langle \theta_{k}-\hat\theta_n, \pk \gk \rangle + \gamma^2 \| \pk \gk \|_2^2 \nonumber\\
      &\le \underbrace{\left( 1 - 2\gamma \underline{\lambda}_P \underline{\lambda}_H  +\gamma^2 [\overline{\lambda}_P  \overline{\lambda}_H]^2 \right)}_{=A(\gamma)}
      \|\theta_{k}-\hat\theta_n\|_2^2,
\end{align*}
where the last inequality is implied by Assumption \ref{ass:Qn} i. and iii. Since  a contraction occurs if $A(\gamma)\in [0,1)$,  global convergence follows. Now at $\gamma=0$,  $A(0)=1$ and $\partial_\gamma A(0)<0$, so  by continuity and local monotonicity of $A(\cdot)$, there exists a nonempty subinterval of the form $(0,\tilde \gamma]$ with $\tilde \gamma \in (0,1]$ such that $A(\gamma) \in [0,1)$ for all $\gamma \in (0,\tilde \gamma]$. This establishes  existence of an interval of values for $\gamma$ close to zero  such that the gradient-based optimizer is globally convergent.
But depending on $\underline{\lambda}_P \underline{\lambda}_H$ and $\overline{\lambda}_P\overline{\lambda}_H$, there may exist larger values of $\gamma \in (0,1]$ with $A(1) \geq 1$ that could frustrate convergence.
The following Lemma shows that $\sqrt{A(\gamma)}$ is the global convergence rate of $\theta_k$ to $\hat\theta_n$.  

\begin{lemma}
\label{lem:cv_non_stochastic}
Suppose Assumption 1 holds, then there exists $\gamma \in (0,1]$ such that $A(\gamma)\in [0,1)$. Let $\overline{\gamma}$ be such that $A(\gamma) = (1-\overline{\gamma})^2$, then
$ \|\theta_k - \hat\theta_n\|_2 \leq (1-\overline{\gamma})^k\|\theta_0-\hat\theta_n\|_2 \to 0, \text{ as } k\to\infty. $
\end{lemma} 
\noindent {\bf Proof of Lemma \ref{lem:cv_non_stochastic}:} As discussed above, there exists  $\gamma$ such that $A(\gamma) \in [0,1)$.   For such $\gamma$, let  $\overline{\gamma}(\underline \lambda_P,\underline \lambda_H,\bar\lambda_K,\bar\lambda_H)  \in (0,1]$  independent of $k$  be such that:
$A(\gamma) = (1-\overline{\gamma})^2  \in [0,1)$. 
It follows that
\begin{align*}
      \|\theta_{k+1}-\hat\theta_n\|_2 &\leq \sqrt{A(\gamma)}\|\theta_{k}-\hat\theta_n\|_2\\& \leq (1-\overline{\gamma})   \|\theta_{k}-\hat\theta_n\|_2 \nonumber \\
      &\leq (1-\overline{\gamma})^{k}   \|\theta_{0} - \hat\theta_n\|_2\rightarrow 0,\quad \text{as } k\rightarrow \infty.\qed
\end{align*}

In general, a larger value of $\overline{\gamma}$  would result in faster convergence of $\|\theta_k-\hat\theta_n\|_2$ to zero. The choice of $\gamma$ and the implied $\overline{\gamma}$ in Lemma \ref{lem:cv_non_stochastic} are typically data-dependent, but further insights can be gained in two special cases. For \gd, the largest globally convergent $\gamma$ is $\underline{\lambda}_H/\overline{\lambda}_H^2$.  In ill-conditionned problems when this ratio is small, convergence will be slow  since $(1-\overline{\gamma})^2 = (1-[\underline{\lambda}_H/\overline{\lambda}_H]^2)$ will be large. For \nr\; when $P(\theta)=H(\theta)^{-1}$,  we can  use $\bar\lambda_{PH}\leq \bar \lambda_{P}\bar \lambda_{H}$ and $\underline \lambda_{PH} \geq \underline\lambda_{P}\underline \lambda_H$ to obtain a tighter bound. The globally convergent $\gamma$ that minimizes $1-2\gamma \underline\lambda_{PH}+\gamma^2 \overline\lambda^2_{PH}$ is then $\gamma  = \underline{\lambda}_{PH}/[\overline{\lambda}_{PH}]^2$ which is strictly less than $1$ for non-quadratic objectives. Since the $(1-\overline{\gamma})^2 = (1-[\underline{\lambda}_{PH}/\overline{\lambda}_{PH}]^2)$ associated with \nr\; is typically smaller than for \gd, \nr\; will converge faster.

\subsection{Consistency of $\overline{\theta}_{\FL}$}
Resampling is usually used for inference, but Algorithm \ref{algo:oe} uses resampling for estimation. Unlike classical optimizers, the resampled gradient is noisy. As a consequence,  the draws $(\thetab)_{b\geq 1}$ constructed by Algorithm \ref{algo:oe} no longer converge deterministically. The following conditions will be imposed on the resampled objective $Q_m^{(b)}$. 
\begin{assumption}
 \label{ass:resampled}
Suppose that $m/n \to c \in [0,1]$ as both $m$ and $n \to +\infty$ and there exists positive and finite constants $C_2,C_3,C_3^\prime,C_4$ such that for all $\theta \in \Theta$, the resampled gradient $G_m^{(b)}(\theta)$ and Hessian $H_m^{(b)}(\theta)$ satisfy the following for all $b\geq 1$ and $\theta\in \Theta$:
      \begin{itemize}
            \item[i.] $\| G_m^{(b)}(\theta) - G_m^{(b)}(\hat \theta_n) - H_m^{(b)}(\hat\theta_n)(\theta-\hat\theta_n) \|_2 \leq C_2  \| \theta-\hat\theta_n\|_2^2$,
            \item[ii.]  $0 < \underline{\lambda}_H \leq \lambda_{\min}(H_m^{(b)}(\theta)) \leq \lambda_{\max}(H_m^{(b)}(\theta)) \leq \overline{\lambda}_H < +\infty$,
            \item[iii.] $\left[ \mathbb{E}^\star \left( \sup_{\theta \in \Theta}\| G_m^{(b)}(\theta)-G_n(\theta)\|_2^2 \right) \right]^{1/2} \leq \frac{C_3}{\sqrt{m}}$,  
\item[iv.]   $\|\mathbb{E}^\star \left( G_m^{(b)}(\hat\theta_n) \right)\|_2  \leq \frac{C_3^\prime}{m}$,
            \item[v.]  $\left[ \mathbb{E}^\star \left( \sup_{\theta \in \Theta}\| H_m^{(b)}(\theta)-H_n(\theta)\|_2^2 \right) \right]^{1/2} \leq \frac{C_4}{\sqrt{m}}$,
            \item[vi.] $0<\underline{\lambda}_P \leq \lambda_{\min}(\pb) \leq \lambda_{\max}(\pb) \leq \overline{\lambda}_P<+\infty$.
      \end{itemize}
\end{assumption}

Assumption \ref{ass:resampled} i. bounds the remainder term in the Taylor expansion of each resampled gradient around the sample minimizer $\hat \theta_n$. Assumption \ref{ass:resampled} ii. implies that each resampled objective is also strongly convex. Conditions iii.-v. are tightness condition on the resampled gradient and Hessian empirical process. It implies uniform convergence over $\Theta$ at a $\sqrt{m}$-rate.\footnote{This is implied by a conditional uniform Central Limit Theorem. See \citet[Chapter 2.9]{VanderVaart1996} and \citet[Chapter 10]{kosorok2007} for iid data. \citet{Chen2003} provide high-level conditions for resampling two-step estimators when the first-step estimator can be nonparametric.}  Condition iv. is satisfied with $C_3^\prime = 0$ for MLE and NLS estimators because $G_n$ is a sample mean. For over-identified GMM, $\overline{g}_n(\hat\theta_n) \neq 0$ and the gradient $G_n(\hat\theta_n) = 2 \partial_\theta \overline{g}_n(\hat\theta_n)^\prime W_n \overline{g}_n(\hat\theta_n)$ is not a sample mean. Condition iv. requires correct specification in GMM so that $\|\overline{g}_n(\hat\theta_n)\|_2$ goes to zero sufficiently fast as $n \to \infty$. 

The following lemma shows that $\theta_b$ will converge in probability to and stays within a $\frac{1}{\sqrt{m}}$ neighborhood of $\hat \theta_n$ as $b$ increases. 

\begin{lemma} \label{lem:cv_stochastic} 
 Under Assumptions \ref{ass:Qn}-\ref{ass:resampled}
and given $\gamma\in(0,1]$ such that $(1-\overline{\gamma})^2=A(\gamma)\in[0,1)$, as defined in Lemma \ref{lem:cv_non_stochastic}, there exists a constant $C_5=C_5(C_3, \overline{\lambda}_P,\gamma)$ such that
      \begin{eqnarray*} 
            \left[ \mathbb{E}^\star \left( \|\theta_{b+1}-\hat\theta_n\|_2^2 \right) \right]^{1/2} 
      &\le &  (1-\overline{\gamma})^{b+1} \bigg[\mathbb{E}^\star(\|\theta_0-\hat\theta_n\|_2^2)\bigg]^{1/2}+\frac{C_5}{\overline{\gamma} \sqrt{m}}  .
      \end{eqnarray*}
\end{lemma}

\noindent {\bf Proof of Lemma \ref{lem:cv_stochastic}:}  
For any $\theta \in \Theta$, let $\scriptgb(\theta) = \sqrt{m} \left( G_m^{(b+1)}(\theta)-G_n(\theta) \right)$.   By construction of $\theta_b$, we have $\thetabf-\hat\theta_n = \thetab-\hat\theta_n-\gamma \pb \gb$.
It follows that
      \begin{eqnarray*}
\thetabf-\hat\theta_n   &=& \thetab-\hat\theta_n-\gamma \pb G_n(\thetab) +\frac{\gamma}{\sqrt{m}} \scriptgb(\thetab).
      \end{eqnarray*}
      Taking the $\|\cdot\|_2$ norm on both sides, applying the triangular inequality and using arguments analogous to Lemma \ref{lem:cv_non_stochastic}, we have for $\gamma \in (0,1]$ small enough such that the same $A(\gamma)\in [0,1)$:
      \begin{eqnarray*}
            \|\thetabf-\hat\theta_n\|_2 &\le& \|\thetab-\hat\theta_n-\gamma \pb G_n(\thetab)\|_2 +\frac{\gamma\overline{\lambda}_P}{\sqrt{m}} \left( \sup_{\theta\in\Theta} \|\scriptgb(\theta)\|_2 \right)\\
            &\le& (1-\overline{\gamma})\| \thetab - \hat\theta_n \|_2 +\frac{\gamma\overline{\lambda}_P}{\sqrt{m}}  \left( \sup_{\theta\in\Theta} \|\scriptgb(\theta)\|_2 \right).
      \end{eqnarray*}
Taking expectations on both sides:
      \begin{eqnarray*}
            \left[ \mathbb{E}^\star \left( \|\thetabf-\hat\theta_n\|_2^2\right)\right]^{1/2}
            &\le& (1-\overline{\gamma}) \left[ \mathbb{E}^\star \left( \| \thetab - \hat\theta_n \|_2^2 \right) \right]^{1/2} +\frac{\gamma\overline{\lambda}_P C_3}{\sqrt{m}}.
      \end{eqnarray*}
 The desired result is then obtained with $C_5 = \gamma\overline{\lambda}_P C_3$.\qed

Lemma \ref{lem:cv_stochastic} shows stochastic convergence of $\theta_b$ to $\hat\theta_n$. To study the properties of our estimator $\bar \theta_{\FL}$, we will use a concept known as {\em coupling}. A coupling between two distributions $\mu$ and $\nu$ on an (unrestricted) common probability space  is a pair of random variables $X$ and $Y$  such that $X\sim \nu$ and $Y \sim \mu$, and  are equal, on average, up to Wasserstein distance of order $p\geq 1$. Precisely,
 the Wasserstein-Fr\'{e}chet-Kantorovich coupling distance  between two distributions $\nu$ and $\mu$ is defined as:
$ W_p(\nu,\mu)^p = \inf_{(X,Y), X \sim \nu, Y \sim \mu} \mathbb{E}(\|X-Y\|^p), p \geq 1. $

Of interest here is the coupling between $\theta_b$ and $\theta_b^\star$, where $\theta_b^\star$ is a linearized sequence of $\theta_b$ defined below. They have different marginal distributions because one is a linear and the other is a  non-linear process. Nonetheless, they live on the same probability space because they rely on the same source of randomness originating from the resampled objective $Q_m^{(b)}$. Hence if we can show that $\|\thetab-\thetab^\star\|$ is small in probability, then we can work with the distribution of $\thetab^\star$ which is more tractable. 

Precisely, we are interested in a  linearized sequence defined as
      \begin{equation} \thetabf^\star - \hat \theta_n = \Psi(\hat\theta_n)(\thetab^\star - \hat\theta_n) - \gamma \bar{P}_m G_m^{(b+1)}(\hat\theta_n),
\label{eq:lin}
\end{equation}
where $\Psi(\hat\theta_n) = I_d- \gamma \bar P_m H_n(\hat\theta_n)$ and $\bar P_m=I_d$ for \rgd\ and $\bar P_m=[H_n]^{-1}$ for \rnr.
We saw earlier from (\ref{eq:rnr-ols}) in the linear regression model that $\Psi(\hat\theta_n)=(1-\gamma)I_d$ for \rnr. We now provide  conditions on $\pb$ for the draws produced in Algorithm \ref{algo:oe} to be close to those defined in (\ref{eq:lin}) in non-quadratic settings.

\begin{assumption} 
\label{ass:K} 
Define $d^2_{0,n}=\mathbb{E}^\star(\|\theta_0-\hat\theta_n\|_2^2)$ and let 
  $\overline{P}_m$ be a symmetric positive definite matrix such that for $\Psi(\hat\theta_n) = I_d- \gamma \bar P_m H_n(\hat\theta)$, 
\begin{itemize}
\item[i.] $0\leq  \lambda_{\max}(\Psi(\hat\theta_n)\Psi(\hat\theta_n)^\prime) < 1$,
\item[ii.] $\left[\mathbb{E}^\star \left( \|I_d-\pb\overline{P}_m ^{-1}\|_2^2 \right) \right]^{1/2}\leq C_6 \left( \rho^b d_{0,n} + \frac{1}{\sqrt{m}} \right)$, for some $\rho \in [0,1)$ and some  $C_6>0$.
 \end{itemize}
\end{assumption}

Assumption \ref{ass:K} ii. is needed to  ensure that the resampled conditioning matrix $\pb$ converges to $\overline{P}_m$ used in (\ref{eq:lin}) and Assumption \ref{ass:K} i. ensures stability of the linearized  process (\ref{eq:lin}). These assumptions  allow us to study $\theta_{b+1}^\star-\hat\theta_n$ as a VAR process with parameters that depend on the Hessian, the conditioning matrix and the learning rate as in the OLS example.

 For \rgd\; with $\pb=\overline{P}_m=I_d$, Condition ii. holds automatically, while Condition i. requires $\gamma<2/\overline{\lambda}_H$.   For \rnr\; with $\overline{P}_m = [H_n(\hat\theta_n)]^{-1}$, it will be shown in Theorem \ref{th:free_lunch}  that  Conditions i.-ii. hold for any $\gamma \in (0,1]$ such that $A(\gamma) \in [0,1)$ under the assumptions of Lemmas \ref{lem:cv_non_stochastic}, \ref{lem:cv_stochastic}.  This implies that  $\thetab^\star$ constructed in (\ref{eq:lin}) for \rnr\; is an AR(1) process with autoregressive coefficient $1-\gamma$ as in the OLS case.  Now define 
\begin{equation}
\label{eq:rhobar}
\overline{\rho} = \max \left[ \sqrt{\lambda_{\max}(\Psi_m(\hat\theta_n)\Psi_m(\hat\theta_n)^\prime)}  ,1-\overline{\gamma},\rho \right] < 1.
\end{equation}
The  autoregressive structure  of  $\theta_b^\star-\hat\theta_n$ and $\thetab-\hat\theta_n$ together with the assumed convergence of $\pb$ to $\overline{P}_m$   lead to the following result on the coupling distance between $\thetab$ and $\theta_b^\star$.

\begin{lemma} \label{lem:coupling}
Suppose that Lemmas \ref{lem:cv_non_stochastic} and \ref{lem:cv_stochastic} hold, and there exists a matrix $\overline{P}_m>0$ satisfying Assumption \ref{ass:K}.  Let $\bar\rho$ be defined as in (\ref{eq:rhobar}). Then
 $\thetab^\star$ defined in (\ref{eq:lin}) satisfies:  
\[ \mathbb{E}^\star \left( \| \thetab - \thetab^\star \|_2 \right) \leq C_7 \left(  \frac{1}{m}+ \overline{\rho}^b[d_{0,n}+d_{0,n}^2] \right). \]
\end{lemma}

 The statement 
 holds for any conditioning matrix $\pb$ evaluated on the subsamples satisfying Assumption \ref{ass:K}. Since  $\sum_{b=1}^B \bar \rho^b\le \frac{1}{1-\overline{\rho}}$,  Lemma \ref{lem:coupling} implies
\begin{align} \mathbb E^\star \bigg(\|\overline{\theta}_{\FL}-\bar \theta^\star_{\FL}\|_2\bigg)\le \frac{C_7 }{1-\overline{\rho}}\bigg(\frac{1}{m}+\frac{d_{0,n}+d_{0,n}^2}{B} \bigg). \label{eq:avg_coupling}
\end{align}

The result is useful because it implies that our estimator $\overline{\theta}_{\FL}$ equals $\overline{\theta}_{\FL}^\star = \frac{1}{B}\sum_{b=1}^B \thetab^\star$ up to vanishing terms. By the triangular inequality:
\begin{align}
      \mathbb{E}^\star\left( \| \overline{\theta}_{\FL} - \hat\theta_n \|_2 \right) &\leq \mathbb{E}^\star\left( \| \overline{\theta}_{\FL} - \overline{\theta}_{\FL}^\star \|_2 \right) + \mathbb{E}^\star\left( \| \overline{\theta}_{\FL}^\star - \hat\theta_n \|_2 \right).
\label{eq:inequality}
\end{align}
The first term can be bounded by Lemma \ref{lem:coupling} as discussed above, and $\mathbb{E}^\star(\thetab^\star) = \hat\theta_n$ by construction of $\thetab^\star$ in (\ref{eq:lin}). Furthermore, Assumption \ref{ass:K} i. implies that the difference $\overline{\theta}_{\FL}^\star - \hat\theta_n$ is a $O_{p^\star}(\frac{1}{\sqrt{mB}})$ since $\thetab^\star$ is asymptotically ergodic and its innovations have variance of order $\frac{1}{m}$.

\begin{theorem}
      \label{th:average_estimation} Let  $\theta^0$ be the population minimizer,  $\hat\theta_n$ be the estimate obtained by a classical optimizer, and  $\{\thetab\}$ be generated by Algorithm \ref{algo:oe}. Suppose that $\{ \sqrt{m}\overline{P}_m G_m^{(b)}(\hat\theta_n) \}_{b \geq 1}$ are iid with finite and bounded variance-covariance matrix. Under the conditions of Lemma \ref{lem:coupling}, 
      \[ \mathbb{E}^\star\left( \| \overline{\theta}_{\FL} - \hat\theta_n \|_2 \right) \leq C_8 \left( \frac{1}{m} + \frac{d_{0,n}+d_{0,n}^2}{B} + \frac{1}{\sqrt{mB}} \right),\]
      where $C_8$ depends on the constants and the largest eigenvalue of $\text{var}^\star(\overline{P}_m G_m^{(b)}(\hat\theta_n))$. Furthermore, suppose that $\frac{\sqrt{n}}{\min(B,m)} \to 0$ and $d_{0,n} = O(1)$ then:
      \[ \sqrt{n} \left( \overline{\theta}_{\FL} - \theta^0 \right) =  \sqrt{n} \left( \hat\theta_n - \theta^0 \right) + o_{p^\star}(1).\]
\end{theorem}
Theorem \ref{th:average_estimation}  says that the average of draws $\overline{\theta}_{\FL}$ is a consistent estimate of $\hat\theta_n$ for any choice of conditioning satisfying Assumption \ref{ass:K}. The inverse Hessian (\rnr) and the identity matrix (\rgd) are examples of such conditioning matrices $\pb$.

\subsection{Asymptotic Validity of \rnr\; for Frequentist Inference}
Theorem \ref{th:average_estimation} is valid for  $\pb$ satisfying the assumptions of the analysis. This subsection specializes to \rnr\; produced by Algorithm \ref{algo:free_lunch}  which uses  the inverse Hessian as  conditioning matrix. There are  two reasons for this choice. First, it implies a faster decline in the initialization bias compared to e.g. $\overline{P}_m = I_d$ used in gradient descent. Second, such a conditioning matrix has a limit  $\overline{P}_m = [H_n(\hat \theta_n)]^{-1}$. For $\overline{P}_m \neq [H_n(\hat \theta_n)]^{-1}$ the dynamics are approximated by a VAR(1) instead of a simple AR(1). While the variance of the AR(1) is proportional to the desired $\mathbb{V}_m$, up to a simple adjustment, this is generally not the case for the VAR(1).

Once Assumption \ref{ass:K} is granted with $\overline{P}_m = [H_n(\hat \theta_n)]^{-1}$, the general idea of deriving the limiting distribution of the \rnr\; draws is to ensure the increasing sum $\thetab^\star = \hat\theta_n - \gamma \sum_{j=0}^{b-1} (1-\gamma)^j H_n(\hat\theta_n)^{-1}G_{m}^{(b-j)}(\hat\theta_n)$ preserves the convergence of each resampled $G_{m}^{(b-j)}(\hat\theta_n)$. The autoregressive nature of $\thetab$ makes the argument somewhat different from the standard setting where each resampled minimizer $\hat\theta_m^{(b)}$ can usually be expressed as a function of a single $G_m^{(b)}(\hat\theta_n)$ plus negligible terms. In such cases, distributional statements about $\hat\theta_m^{(b)}$ follow from the convergence of each resampled $G_m^{(b)}$. Here, the increasing sum $\thetab^\star$ depends on the entire history of the independently resampled $\{G_{m}^{(b-j)}(\hat\theta_n)\}_{j=0,\dots,b-1}$ for which we need to prove convergence.

\begin{assumption} \label{ass:higher_order} Let $\overline{P}_m$ in Theorem \ref{th:average_estimation} be $[H_n(\hat\theta_n)]^{-1}$. Suppose that $\{ \sqrt{m}\overline{P}_m G_m^{(b)}(\hat\theta_n) \}_{b \geq 1}$  has  a non-singular variance-covariance matrix denoted $\mathbb{V}_m$.  For some $\beta \in (0,1/2]$ and $\|r_m(\tau)\| \leq C_\psi \|\tau\|^{\alpha}$ with $\alpha > 0$, it holds that for $\mathbf{i}^2=-1$:
\[ \mathbb{E}^\star \left(  \exp \left[\sqrt{m} \mathbf{i}\tau^\prime (\mathbb{V}_m)^{-1/2} [H_n(\hat\theta_n)]^{-1}G_m^{(b)}(\hat\theta_n) \right] \right) = \exp \left( - \frac{\|\tau\|_2^2}{2}\right) \cdot \left( 1+ \frac{r_m(\tau)}{m^{\beta}} \right).\]
\end{assumption}

Assumption \ref{ass:higher_order} requires non-degeneracy of the variance-covariance matrix which is required for Central Limit Theorems \citep[][Theorem 5.3]{white2014}.
 Assumption \ref{ass:higher_order} provides higher-order conditions to ensure that the bootstrap converges in distribution at a sufficiently fast rate. It can be understood as requiring  the resampled data to have an Edgeworth expansion, the first term being the characteristic function of the standard normal distribution. This  occurs with $\beta=1/2$, $\alpha=1$ for averages of iid data with finite third moment \citep[][Chapters 6.2-6.3]{lahiri2006}. By Assumption \ref{ass:higher_order}, the error in the Gaussian approximation of $\sqrt{n}[H_n(\hat\theta_n)]^{-1}G_m^{(b)}(\hat\theta_n)$ depends on $\alpha$ through $r_m(\tau)$ and on $\beta$ through the inflation factor $1+\frac{r_m(\tau)}{m^\beta}$. These two parameters are of significance because the error in the  asymptotic approximation for $\thetab^\star$ inherits the error in $\sqrt{n}[H_n(\hat\theta_n)]^{-1} G_m^{(b)}$. The following theorem takes as given the validity of bootstrap standard errors, i.e. $\sqrt{m} (\mathbb{V}_m)^{-1/2} \Big( \hat\theta_n - \theta^0 \Big) \overset{d}{\to} \mathcal{N}\left( 0, I_d \right)$.
\begin{theorem}  \label{th:free_lunch} Let $\{\thetab\}$ be generated by Algorithm \ref{algo:free_lunch} and suppose that the conditions of Lemmas \ref{lem:cv_non_stochastic}, \ref{lem:cv_stochastic} hold then Assumption \ref{ass:K} holds with $\overline{P}_m = [H_n(\hat\theta_n)]^{-1}$. Furthermore, suppose Assumption \ref{ass:higher_order} holds and let $\phi(\gamma)=\frac{\gamma^2}{1-(1-\gamma)^2}$, then as $m,b \to +\infty$ with $\log(m)/b \to 0$,
\[ (\phi(\gamma)\mathbb{V}_m)^{-1/2}\sqrt{m} \left( \thetab - \hat\theta_n \right) \overset{d^\star}{\to} \mathcal{N}\left( 0, I_d \right).\]
\end{theorem}

The thrust of the Theorem is that $[H_n(\hat\theta_n)]^{-1} G_m^{(b)}(\hat\theta_n)$ is approximately normal when properly standardized by $(\mathbb{V}_m)^{-1/2}$ and scaled by $\sqrt{m}$. The summation in the AR(1) representation (\ref{eq:AR1_OLS}) preserves this property 
under the stated assumption but inflates the variance by a factor $\phi(\gamma)$ which needs to be adjusted. As pointed out above, the error in the Gaussian approximation of $\thetab^\star$ is of the same order as $\sqrt{n}[H_n(\hat\theta_n)]^{-1}G_m^{(b)}(\hat\theta_n)$, which depends on  $\alpha,\beta$ according to Assumption \ref{ass:higher_order}. But $r_m(\tau)$ is inflated by a factor of $\frac{(2-\gamma)^\alpha}{1-[1-\gamma]^\alpha}$ which is $1$ when $\gamma=1$ and goes to infinity as $\gamma \to 0$. The Gaussian approximation is better for larger $\gamma$.

An implication of Theorems \ref{th:average_estimation} and \ref{th:free_lunch} is that
\[ \mathbb V_{\rnr}^{-1/2}\sqrt{n}\left( \overline{\theta}_{\rnr} - \theta^0 \right) = \mathbb V_{\rnr}^{-1/2}\sqrt{n}\left( \hat\theta_n - \theta^0 \right) + o_{p^\star}(1) \overset{d}{\to} \mathcal{N}\left( 0, I_d \right), \]
where $\mathbb{V}_{\rnr} =  \frac{m}{\phi(\gamma)}\text{var}^\star(\thetab-\hat\theta_n)$, and a plug-in estimator $V_{\rnr}$ is defined in Algorithm \ref{algo:free_lunch}.
This implies that  standard errors and quantiles computed from the draws $\thetab$, after adjusting for $m$ and $\phi(\gamma)$, can be used to make asymptotically valid inference. Confidence intervals can be constructed to test linear and non-linear hypotheses. 



Theorem \ref{th:free_lunch} specializes to \rnr\; where $P_b=[H_m^{(b+1)}(\thetab)]^{-1}$. Because the AR(1) representation in (\ref{eq:lin}) does not hold for \rgd,   simple adjustments cannot be designed that would  allow \rgd\; to provide valid inference. Furthermore, when $\gamma \in (0,1]$ is small enough such that \rgd\, converges, it is not uncommon that  $\lambda_{\max}(\Psi(\hat\theta_n)\Psi(\hat\theta_n)^\prime) \simeq 1$ because of ill-conditioning, and when $\thetab$ is very persistent,  a much larger $B$ will be required.

Given the Markov chain nature of our $\thetab$, convergence of the chain can be diagnosed using the standard tools from the MCMC literature such as  convergence diagnostics considered in \citet{gelman1992,Gelman1998}. As seen from (\ref{eq:lin}), the draws $\thetab$ approximately follow $d$ univariate AR(1) processes with the same persistence parameter $(1-\gamma)$ which is user-chosen. This can be used to gauge the quality of our large sample approximation in the data for a given pair $(\gamma,m)$. We will illustrate this feature below.

It is noteworthy that while the appeal of stochastic optimization is the savings from using $m \ll n$, our \rnr\;  requires $m$ not to be too small. This can be seen as the cost of valid inference. Nonetheless, several additional shortcuts could improve the numerical performance of \rnr. Our algorithm can be modified so that the Hessian is updated every few iterations rather than at each iteration.  The draws would still be valid since the assumptions of Theorem \ref{th:free_lunch} would still hold. The Hessian could also be approximated using quasi-Newton methods which only requires computing gradients.  However, as shown in \citet{dennis1977}, \citet{nocedal-wright:06}, the analytical properties of the Hessian approximated by \textsc{bfgs} can only be guaranteed under strong conditions for quadratic objectives. \citet{ren1983} show that  the \textsc{bfgs} estimate $P_k$ may not converge to the Hessian even for quadratic objectives. Theoretical guarantees can be given for less popular but more tractable methods such as Broyden's method or the Symmetric Rank-1 (SR1) update \citep{Conn1991}. These, unfortunately, tend to be less stable than \textsc{bfgs} even in classical optimization. Though an extension of Theorems \ref{th:average_estimation} and  \ref{algo:free_lunch} to quasi-Newton updating is left to future work, the results  based on a resampled \textsc{bfgs}  procedure are promising, as will be seen below.

\subsection{Relation to other Bootstrap and Quasi-Bayes Methods} \label{sec:boot}
Our algorithm is related to several other fast bootstrap methods.
 As discussed in the introduction, solving the minimization problem $B$ times can be computationally challenging or infeasible. Some shortcuts have been proposed to generate bootstrap draws for inference at a lower cost.  \citet{Davidson1999} (hereafter  \dmk) proposes a $n$ out of $n$ {\em approximate} bootstrap that replaces non-linear estimation on each batch of re-sampled data by a small number of Newton steps using $\hat\theta_n$ as starting value.
 In our notation, they perform  Newton-Raphson updating 
$\theta^{(b)}_{\dmk,j+1} = \theta^{(b)}_{\dmk,j} - [H_n^{(b)}(\theta^{(b)}_{\dmk,j})]^{-1} G_n^{(b)}(\theta^{(b)}_{\dmk,j})$ with $\theta^{(b)}_{\dmk,0} = \hat\theta_n$ and $j=0,\dots,k-1$ times for each $b=1,\ldots, B$ and report the draws $\theta^{(b)}_{\dmk,k}$. \citet{Armstrong2014} extends this approach for two-step estimation with a finite dimensional or nonparametric first-step estimator.
\citet{Kline2012} (hereafter,  \ks) suggests a {\em score bootstrap} that uses random weights to perturb the score  while holding the Hessian at the sample estimate.  If the random weights are $\{ \omega^{(b)}_i\}$ with $\mathbb E[ \omega_i]=0, \mathbb E[\omega_i^2]=1$,  then the distribution $[H_n(\hat\theta_n)]^{-1}\frac{1}{\sqrt{n}} \sum_{i=1}^n  \omega^{(b)}_i G_i(\hat\theta_n; y_i,x_i)$ conditional on the data is used to approximate that of $\sqrt{n}(\hat\theta_n-\theta^0)$. 
The appeal is that  the Hessian only needs to be computed once.  \citet{Honore2017} proposes an approach where the resampled objective is minimized only in a scalar direction for a class of models.

The methods above all rely on a preliminary converged estimate, $\hat\theta_n$ and hence estimation precedes inference.   We compute   $\bar\theta_{\rnr}$ and an estimate of its sampling uncertainty in the same loop, so no further computation is needed once $\bar\theta_{\rnr}$ is available. Under our assumptions, the initialization bias will vanish. The practical implication is that for $B$ large enough, the initial values of  \rnr\ can be far  away from the global minimum $\hat \theta_n$, and the algorithm  will   not be sensitive to the usual stopping criteria used in optimization to find $\hat \theta_n$. 

\citet{liang-su:19}  suggests a `moment-adjusted'  algorithm (\textsc{masgrad}) that,  in our notation, updates according to $\gamma_b \to 0$ with $\pb = \text{var}(\sqrt{n}G_n(\theta))^{-1/2}$, which is the asymptotic variance-covariance matrix of the sample gradient. In practice, they recommend to evaluate this quantity using the full sample. Under the information matrix equality, we have  $\mathbb E[(G_n(\thetab) G_n(\thetab)']= \mathbb E [H(\thetab)]$ so that the difference  amounts to using $H(\thetab)^{-1/2}$ instead of $H(\thetab)^{-1}$. While such a conditioning matrix would result in consistent estimates, it would not provide asymptotically valid bootstrap draws,  which requires $\pb =  H(\thetab)^{-1}$ and $\gamma$ fixed as shown in our Theorem \ref{th:free_lunch}. 

The \textsc{sgld} algorithm proposed in \citet{welling-teh:11} updates according to
\begin{equation}
\label{eq:sgld}  \thetabf=\thetab+ \frac{\gamma_{b+1}}{2}   \bigg(\nabla\log p(\thetab)+ \frac{n}{m}\sum_{i=1}^m   \nabla\log p(x^{(b)}_i|\thetab) \bigg) + v_{b+1}
\end{equation}
where $v_{b}\sim N(0,\gamma_b I_d)$ is an injected noise, $\gamma_b$ satisfies (\ref{eq:munro-robbins-conditions}) and $p(\thetab)$ is the prior distribution evaluated at $\thetab$ while $\log p(x^{(b)}_i|\thetab)$ is the log-likelihood of a resampled observation $x^{(b)}_i$ evalutated a $\thetab$.  The update is thus based on the gradient of the log posterior distribution. Like \textsc{sgld} draws, the draws of our free-lunch bootstrap involve two phases: optimization and sampling. First, in the optimization phase, the shape of the objective function dominates the resampling noise until $\thetab$ attains a neighborhood of $\hat\theta_n$. Then, in the sampling the resampling phase, the noise dominates and \rnr\, draws have bootstrap properties.  Compared to \textsc{sgld}, the noise is not injected exogenously and our $\gamma$ is fixed.  \citet{welling-teh:11} shows that with carefully chosen step size $\gamma_b$ and noise variance $\sigma^2_v$, \textsc{sgld} draws can be used for Bayesian inference. Our free-lunch algorithm  does not involve any prior and the goal is frequentist inference, as in the Laplace-type inference proposed in \citet{Chernozhukov2003} (hereafter CH).

 Like CH, our goal is also to simplify the estimation of complex models. CH tackles non-smooth and non-convex objective functions by combining a prior with a  transformation of the objective function.   In principle, we can also handle non-convex objective functions through regularization, but smoothness is an assumption we need to maintain. CH relies on a Laplace approximation to validate the theory while we use the idea of coupling. By nature of the Metropolis-Hastings algorithm, not all CH draws are accepted and the Markov chain is better described as a threshold autoregressive process. All our draws are accepted and they constitute a  nonlinear but smooth autoregressive process. Valid quasi-Bayes inference  requires  the optimal weighting matrix $W_n=\var(\sqrt{n}(\bar g_n(\hat\theta_n))$ which  needs to  be estimated. Continuously updating $W_n(\theta)$ can result in local optima so that the MCMC chain can take significantly more time to converge. Whether or not convexification is required, our approach does not require a specific weighting matrix. 

In terms of tuning parameters,  CH requires as input the proposal distribution in the Metropolis-Hastings algorithm and the associated hyper-parameters. Our tuning parameters are confined to the fixed learning rate $\gamma$ and the resampling size $m$, which do not depend on the dimension of $\theta$. The complexity of the problem also affects the two algorithms in different ways.  As seen from Lemma \ref{lem:cv_non_stochastic}, \nr\; converges at a dimension-free linear rate of $(1-\overline{\gamma})$, whereas  MCMC converges more slowly as the dimension of $\theta$ increases. For instance, the number of iterations needed for the random walk Metropolis-Hastings to converge increases quadratically with the condition number of the Hessian of the log-density and linearly in the dimension $d$ of $\theta$. To alleviate this issue, several samplers exploit gradient information  \citep{roberts1996,girolami2011,neal2011,welling-teh:11}. While these methods improve upon random walk Metropolis-Hastings, ill-conditioning can still render slow convergence. Scaling the proposal using Hessian information can reduce the effect of ill-conditioning but requires a preliminary estimate. See \citet{dwivedi2019log}, Table 1, for an overview of mixing times in Metropolis-Hastings algorithms.

\section{Examples} \label{sec:examples}

This section illustrates the properties of the \rnr\; draws using simulated data and data used in published work.  Throughout, we use a burn-in period of $\textsc{burn} = 1+\text{round}( \log(0.01)/\log(1-\gamma))$ so that the bias is approximately less than 1\% of the initialization error $\|\theta_0-\hat\theta_n\|$. Additional implentation details are given in Appendix \ref{apx:additional}. The set of  $\gamma$ values  satisfying the conditions for Lemma \ref{lem:cv_non_stochastic} are data dependent, but  in all simulated and empirical examples, $\gamma \in [0.1,0.3]$ performed well.

 
\subsection{Simulated Examples}
\paragraph{Example 1: OLS}
 We simulate data from the linear model with intercept $\beta_0=1$, slope $\beta_1=1$,
$x_i \sim \mathcal E(2)$, $e_i \sim t(6)$, $n=200$. We set $B=1000$ plus burn-in draws.
 Homoskedastic standard errors with a degree of freedom adjustment are computed. Table \ref{tbl:table-ols} reports estimates and standard errors for one simulated sample. We consider three values of batch size $m=200,50,10$ and for each batch size, three values of the learning rate $\gamma$. The results are denoted \rnr$_\gamma$ for  $\gamma=1, 0.1$, and $0.01$. The smaller the $\gamma$, the more persistent are the draws. Thus $\gamma=0.01$ is  a  case of extreme  persistence, and as seen from the analysis of the linear model, the variance of the draws are larger the smaller $\gamma$ is. 

\begin{table}[H] \caption{OLS: Estimates and Standard Errors for $\beta_1$}
  \centering
\label{tbl:table-ols}
  \begin{tabular}{l|baaa|bbaaa}
    \hline \hline
    & \multicolumn{4}{c|}{Estimates} & \multicolumn{5}{c}{Standard Errors}\\
   m & \mc{1}{\textsc{ols}} & \mc{1}{$\rnr_1$} & \mc{1}{$\rnr_{0.1}$} & \multicolumn{1}{c|}{$\rnr_{0.01}$} & \mc{1}{\textsc{ase}} & \mc{1}{\textsc{boot}} & \mc{1}{$\rnr_1$} & \mc{1}{$\rnr_{0.1}$} & \mc{1}{$\rnr_{0.01}$} \\ 
    \hline
    200 & 1.230 & 1.236 & 1.234 & 1.234 & 0.180 & 0.159 & 0.164 & 0.155 & 0.193 \\ 
    50 & - & 1.251 & 1.241 & 1.262 & - & 0.184 & 0.179 & 0.187 & 0.161 \\ 
    10 & - & 1.288 & 1.258 & 1.296 & - & 0.255 & 0.270 & 0.254 & 0.205 \\
     \hline \hline
  \end{tabular}\\
  \textit{Remark: Results reported for one simulated sample of size $n=200$.}
  \end{table}
The OLS estimator takes the value $\hat \beta_1 = 1.230$ for this simulated sample. 
We see from Table \ref{tbl:table-ols} that the \rnr\, estimate is very close to OLS when $m=200$ ($=n$) and the choice of $\gamma$ makes little difference.   Theorem \ref{th:average_estimation} suggests that the estimation error should be of order $\frac{1}{\sqrt{m}}$. The large bias associated with a  $m$ small  is most visible at $m=10$, which is less than $\sqrt{n}$.  The difference between the OLS and \rnr\, estimates is nearly a third of a standard error for $\gamma=1$. The $m$ out of $n$ Bootstrap and \rnr\; standard errors are also less accurate with $m=10$. 


\paragraph{ Example 2: MA(1)}
Consider the estimation of a MA(1) model by non-linear least squares (\textsc{nlls}).
 The data is generated as
$y_t = \mu +  e_t + \psi e_{t-1} $.
We set $\mu = 0, \psi = 0.8$, $n = 500$ and $B=2,000$. In this example, $ Q_n(\theta) = \sum_{t=1}^n e_t(\theta)^2 $
where $e_t (\theta)$ are the \textsc{nlls} filtered residuals computed as described in Appendix \ref{apx:additional}. In estimation, the gradient and Hessian are computed analytically. For the standard bootstrap, we implement a state-space resampling algorithm described in Appendix \ref{apx:additional}. 
For \rnr, we initialize at $\theta_0 = (0,0)$ with a learning rate set to $\gamma=0.6,0.1$ and $0.01$, noting that $\gamma=1$ was too large to get stable results in this example. 

\begin{table}[H] \caption{MA(1): Estimates of $\psi$ and Standard Errors}
\label{tbl:table-MA1}
  \centering
  \begin{tabular}{l|baaa|bbbaaa}
    \hline \hline
    & \multicolumn{4}{c|}{Estimates} & \multicolumn{6}{c}{Standard Errors} \\
  $m$ & \mc{1}{\textsc{nlls}} & \mc{1}{$\rnr_{0.6}$} & \mc{1}{$\rnr_{0.1}$} & \multicolumn{1}{c|}{$\rnr_{0.01}$} & \mc{1}{\textsc{ase}} & \mc{1}{\textsc{boot}} & \mc{1}{\dmk} &  \mc{1}{$\rnr_{0.6}$} & \mc{1}{$\rnr_{0.1}$} & \mc{1}{$\rnr_{0.01}$} \\ 
    \hline
    500 & 0.816 & 0.825 & 0.822 & 0.820 & 0.026 & 0.027 & 0.023 & 0.025 & 0.029 & 0.113 \\ 
    250 & - & 0.819 & 0.819 & 0.814 & - & 0.028 & - & 0.034 & 0.034 & 0.081 \\ 
    50 & - & 0.805 & 0.786 & 0.780 & - & 0.035 & - & 0.042 & 0.040 & 0.050 \\ 
     \hline \hline
  \end{tabular}\\
  \textit{Remark: Results reported for one simulated sample of size $n=500$.}
  \end{table}

In this synthetic data, the \textsc{nlls} estimator is $\hat \psi = 0.816$.
Table \ref{tbl:table-MA1} shows that when $m=n$, \rnr\; produces a $\overline{\theta}_{\rnr}$ that is very close to the full sample \textsc{nlls} estimate for all three values of $\gamma$. As in the OLS example  above, the bias and  standard errors are larger when $m$ is smaller, as suggested by Theorem \ref{th:average_estimation}. The \rnr\, standard errors are very similar to those obtained by the  $m$ out of $n$ bootstrap for all values of $m$. The standard errors are quite poor for $\gamma=0.01$, most likely because of the strong persistence of the draws.




\subsection{Empirical Examples}
This subsection considers three examples, the first concerns probit estimation of labor force participation, the second is covariance structure estimation of earnings dynamics, and the third is structural estimation of a BLP model. 
\paragraph{Application 1: Labor Force Participation}
The probit model is of interest because the objective function is strictly convex.
To illustrate, we estimate the model for female labor force participation considered in \citet{mroz:87}.  The data consist of $n=753$ observations assumed iid. We set $B=1000$ and $\gamma=0.3$. Three values of $m$ are considered: $m=n$, $200$, $100$. Appendix \ref{apx:Rcode} provides \textsc{r} code for replicating \rnr\; in this example. There are 8 parameters in this exercise and to conserve space, we only report 4 to get a flavor of the results. Table \ref{tbl:table-mroz_apx} in the on-line Appendix reports all coefficients. As seen from  Table \ref{tbl:table-mroz}, the \rnr\; estimates are close to the MLE ones. Furthermore, the \rnr\; standard errors are close to the bootstrap standard errors. Table \ref{tbl:table-mroz} also shows results for resampled \textsc{bfgs} which is labeled r\textsc{qn}. Evidently, the r\textsc{qn} estimates are similar to \rnr; but is much faster to compute because the Hessian is not computed directly.  
 


\begin{table}[ht]
      \centering  \caption{Labor Force Participation: Estimates and Standard Errors} \label{tbl:table-mroz} \setlength\tabcolsep{5.0pt} \centering
      \begin{tabular}{l|bbbbaaaaaa}
            \hline \hline 
            & \multicolumn{10}{c}{Estimates}\\
       & \textsc{mle}  &  &  &  &  \mc{1}{\rnr$_n$} & \mc{1}{\rnr$_{200}$} & \mc{1}{\rnr$_{100}$} & \mc{1}{r\textsc{qn}$_n$} & \mc{1}{r\textsc{qn}$_{200}$} & \multicolumn{1}{c}{r\textsc{qn}$_{100}$}  \\ 
        \hline
        nwifeinc & -0.012 & - & - & - & -0.012 & -0.013 & -0.014 & -0.012 & -0.011 & -0.012 \\ 
  educ & 0.131 & - & - & - & 0.132 & 0.138 & 0.143 & 0.131 & 0.129 & 0.129 \\ 
  exper & 0.123 & - & - & - & 0.123 & 0.124 & 0.123 & 0.123 & 0.124 & 0.125 \\ 
  exper2 & -0.002 & - & - & - & -0.002 & -0.002 & -0.002 & -0.002 & -0.002 & -0.002 \\
         \hline
         & \multicolumn{10}{c}{Standard Errors} \\
         & \textsc{ase} & \textsc{boot} & \dmk & \ks & \mc{1}{\rnr$_n$} & \mc{1}{\rnr$_{200}$} & \mc{1}{\rnr$_{100}$} & \mc{1}{r\textsc{qn}$_n$} & \mc{1}{r\textsc{qn}$_{200}$} & \multicolumn{1}{c}{r\textsc{qn}$_{100}$}\\  \hline
         nwifeinc & 0.005 & 0.005 & 0.005 & 0.005 & 0.005 & 0.006 & 0.005 & 0.005 & 0.005 & 0.005 \\ 
  educ & 0.025 & 0.026 & 0.026 & 0.025 & 0.025 & 0.027 & 0.028 & 0.027 & 0.025 & 0.025 \\ 
  exper & 0.019 & 0.020 & 0.019 & 0.019 & 0.019 & 0.020 & 0.021 & 0.019 & 0.018 & 0.017 \\ 
  exper2 & 0.001 & 0.001 & 0.001 & 0.001 & 0.001 & 0.001 & 0.001 & 0.001 & 0.001 & 0.001 \\ 
        \hline  \hline
      \end{tabular}
\end{table}
      
\begin{figure}[H] \caption{Labor force participation: draws for $\theta_{\text{educ}}$} \label{fig:mroz}
      \centering  \vspace{0.2cm}
      \includegraphics[scale=1.1]{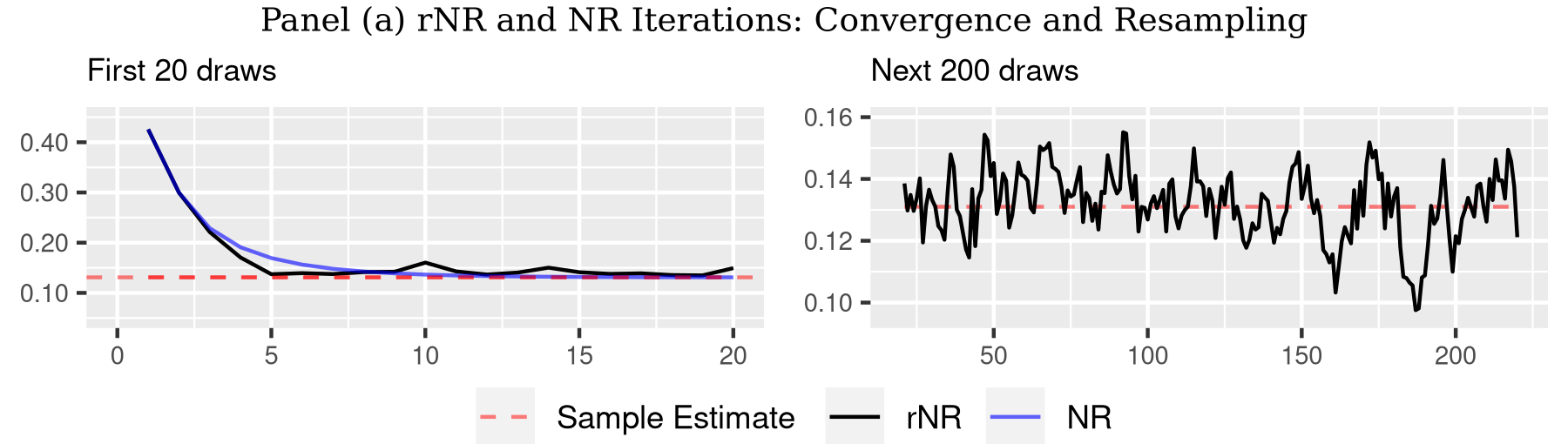}\\ \vspace{0.2cm}
      \includegraphics[scale=1.1]{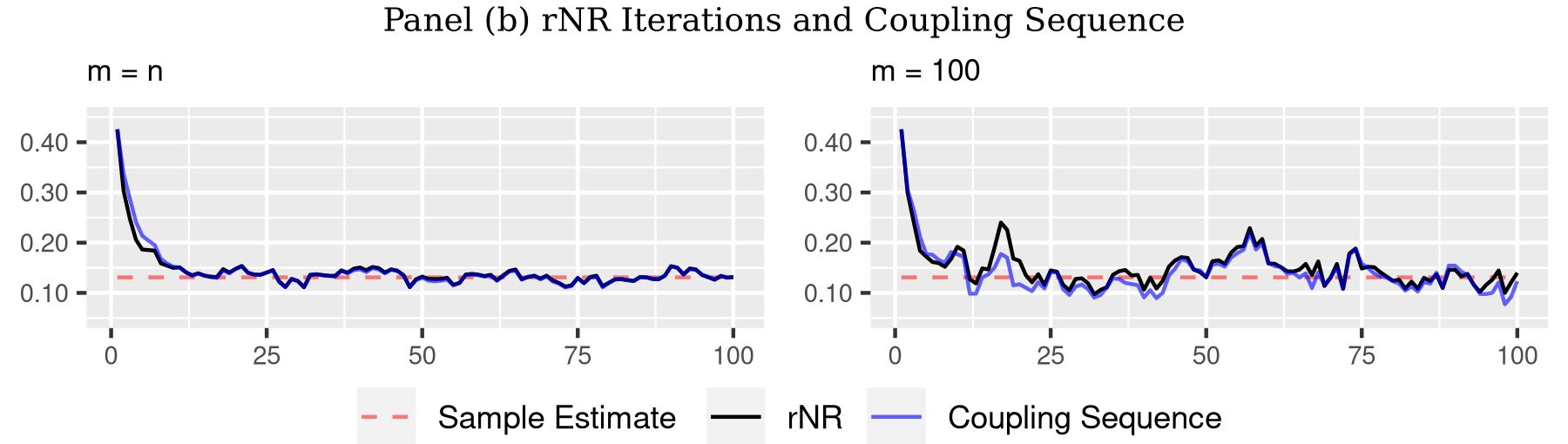}\\ \vspace{0.2cm}
      \includegraphics[scale=1.1]{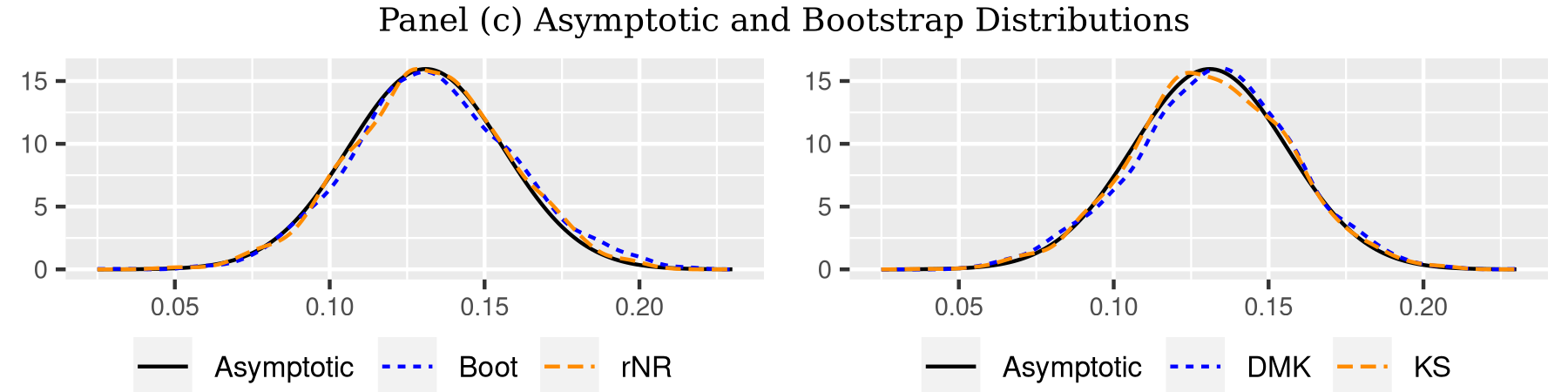}
    \end{figure}

Panel (a) of Figure \ref{fig:mroz} illustrates the behavior of the draws produced by \rnr. The dashed red line corresponds to the MLE estimate $\hat\theta_n$. The black line corresponds to \rnr\; draws based on resampling the data with replacement. The blue line shows iterates from classical \nr\; with the same $\gamma=0.3$. The top left panel shows the first $20$ draws  in the  convergence phase when the classical \nr\;  and the proposed \rnr\; should  behave  similarly. While in this example,  \rnr\; converges after  $5$ draws, \textsc{nr} requires  $10$ to $15$ iterations to achieve convergence. The top right panel plots the next $200$ draws. Since convergence is achieved after 5 draws, these draws are in  the re-sampling phase.   Evidently, the transition between the convergence and the resampling phase of \rnr\; is seamless.  The AR(1) coefficient on $\theta_{b,educ}$ based on  the converged draws (after discarding the first five) is estimated to be  $0.673$ with a  standard error $0.016$, which is not significantly different from $0.7=1-\gamma$ predicted by Lemma \ref{lem:coupling}.

Panel (b) of Figure \ref{fig:mroz} uses the \citet{mroz:87} data to further illustrate Lemma \ref{lem:coupling}.  We compare the \rnr\; draws with two  AR(1) sequences generated according to  coupling theory in (\ref{eq:lin}), ie.   $\thetabf^\star = \hat\theta_n + (1-\gamma)(\thetab^\star-\hat\theta_n) - \gamma H_n^{-1}G_m^{(b+1)}(\hat\theta_n)$ with  $\theta^\star_0 = \theta_0$.  For $m=n$ shown in the left panel, the coupling result is very accurate after the short initial convergence phase as the two series are nearly indistinguishable. The right panel shows that coupling distance is noticeably greater when $m=100$.

Panel (c) of Figure \ref{fig:mroz} illustrates Theorem \ref{th:free_lunch} by comparing the asymptotic Gaussian distribution with the bootstrap, \rnr, \dmk\, and \ks\, distributions for the education coefficient with $m=n$ and $\gamma=0.3$. The distribution of the \rnr\, draws is rescaled using the simple adjustment: $\overline{\theta}_{\rnr} + \sqrt{\frac{m}{n\phi(\gamma)}}(\thetab-\overline{\theta}_{\rnr})$ after discarding a burn-in period of $10$ draws. The \rnr\, distribution approximates the bootstrap distribution quite well. 

\paragraph{Application 2:  Earnings Dynamics}

\citet{moffitt-zhang:18} estimates  earnings volatility using a subsample of $3508$ males in the Panel Study of Income Dynamics (PSID) dataset between 1970 and 2014 for a total of $36403$ observations. Let $y_{iat}$ denote an individual's earnings $i$ in age group $a$ (between 24 and 54) at time $t$.  Earnings are assumed to be the sum of a permanent $\mu_{ia}$ and a transitory  $\nu_{iat}$ component:
\begin{align*}
  y_{iat}  = \alpha_t \mu_{ia} + \beta_t \nu_{ia}, \quad
  \mu_{ia} = \mu_{i0} + \sum_{s=1}^a \omega_{is}, \quad
  \nu_{ia} = \varepsilon_{ia} + \sum_{s=1}^a \psi_{a,a-s} \varepsilon_{is} \text{, for } a \geq 2.
\end{align*}
In \citet{moffitt-zhang:18}, the variances are modeled via 11 parameters and estimated by a sequential quadratic programming algorithm (SQP). The Hessian in their example  has both positive and negative eigenvalues, suggesting that the solution could be a saddle point.  To abstract from identification issues,  we estimate $\theta=(\nu_0,\delta_0,\gamma_0,\gamma_1)$ and fix the remaining 7 parameters.\footnote{Specifically,
$  \text{var}(\mu_{i,0})$  by $\nu_0$,
$  \text{var}(\omega_{ir})$  by $\delta_0, \delta_1$,
$  \text{var}(\varepsilon_{ir})$ by $ \gamma_0,\gamma_1,k$,
and $  \psi_{a,a-r}$ by $ \pi, \lambda_1, \eta_1,\eta_2,\eta_3$. We set $k=1,\lambda=5$, and all remaining parameters to zero.} We specify the conditioning matrix as $\pb = (\hb^\prime\hb)^{-1/2}$ to ensure positive definiteness. Algorithm \ref{algo:free_lunch} converges  using  the starting values   $\theta_0 = ( 0.054, -10.257, -4.355, 0.012 )$ as in the original paper, with $m=n$, $\gamma=0.2$  $B=2000$, and resampling at the age-cohort level. We also consider re-weighting instead of resampling which is denoted as \rnr$_w$. 
Though our theory does not cover resampled quasi-Newton methods,  we also use an implementation of \textsc{bfgs} that sets  $\pb = (H_{b,\textsc{bfgs}}^\prime H_{b,\textsc{bfgs}})^{-1/2}$, where $H_{b,\textsc{bfgs}}$ is the \textsc{bfgs} approximation of the Hessian matrix, and report the results as r\textsc{qn}.  

Table \ref{tbl:table-earnings} shows that the \rnr, \rnr$_w$ and r\textsc{qn} estimates are very close to $\hat\theta_n$ obtained by SQP. The \rnr$_w$\; standard errors are larger than the \rnr\; ones, which are in turn larger than the bootstrap ones, but the differences are not enough to change the conclusion that all four parameters are statistically  different from zero. However, bootstrap inference of $\hat\theta_n$ is  time-consuming,  requiring 5h48m to produce 2000 draws, even after the original Matlab code was ported to \textsc{r} and C++ using Rcpp to get a greater than $10$ times  speedup in computation time.  In contrast,  \rnr\; produces estimates and standard errors  in  1h4m, and the r\textsc{qn}  in 38m. The time needed for \dmk\  to produce standard errors is comparable to \rnr, while r\textsc{qn} is more comparable to \ks. However, both have an overhead of having to first obtain $\hat\theta_n$, which entails minimization of the objective function by SQP. 


\begin{table}[ht] \caption{Earnings Volatility: Estimates and Standard Errors} \label{tbl:table-earnings}
      \centering \setlength\tabcolsep{4.15pt}
      \begin{tabular}{l|baaaa|bbbaaaa}
            \hline \hline 
            & \multicolumn{5}{c|}{Estimates} & \multicolumn{7}{c}{Standard Errors} \\
        & \mc{1}{$\hat\theta_n$} & \mc{1}{\rnr} & \multicolumn{1}{c}{\rnr$_w$} & \multicolumn{1}{c}{r\textsc{qn}} & \multicolumn{1}{c|}{r\textsc{qn}$_w$} & \textsc{boot} &\textsc{dmk} & \textsc{ks} & \mc{1}{\rnr} & \mc{1}{\rnr$_w$} & \mc{1}{r\textsc{qn}} & \multicolumn{1}{c}{r\textsc{qn}$_w$} \\ \hline
        $\nu_0$  & 0.109 & 0.109 & 0.109 & 0.109 & 0.109 & 0.002 & 0.002 & 0.002 & 0.002 & 0.002 & 0.002 & 0.002 \\ 
        $\delta_0$ & -5.768 & -5.779 & -5.779 & -5.767 & -5.766 & 0.050 & 0.062 & 0.049 & 0.063 & 0.060 & 0.051 & 0.049 \\ 
        $\gamma_0$ & -1.839 & -1.819 & -1.819 & -1.841 & -1.842 & 0.083 & 0.101 & 0.079 & 0.094 & 0.091 & 0.089 & 0.082 \\ 
        $\gamma_1$ & 0.010 & 0.011 & 0.011 & 0.010 & 0.010 & 0.010 & 0.012 & 0.010 & 0.011 & 0.012 & 0.011 & 0.011 \\
         \hline
           & \mc{1}{}& \mc{1}{} & \mc{1}{} & \mc{1}{} & \multicolumn{1}{c|}{time} & 5h48m & 1h4m & 13m & 1h4m & 1h4m  & 38m & 38m\\ \hline\hline
      \end{tabular}
\end{table}

 In addition to providing standard errors, an additional by-product of  Algorithm \ref{algo:free_lunch} is that the draws can be used for model diagnostics. \citet{Gentzkow2017} provides statistics to assess  the sensitivity  of  the parameter estimates to assumptions of the model, taking the data as given. Our algorithm takes the model assumptions as given, but takes advantage of  resampling  to shed light on the  sensitivity of the estimates to features of the data themselves.  As pointed out in \citet{Chatterjee1986}, influential observations could be outliers, or  could be  points of high leverage.   If no such observations exist, removing them in the resampled data should not significantly affect the Markov chain. If their presence is influential, we should witness a `break' in the draws. 

With this motivation in mind, we examine whether the parameter estimates of the earnings model are sensitive to data of a particular age group.
  \begin{figure}[ht]
\caption{Earnings Volatility: Sensitivity to Age Groups} \label{fig:moffitt}
\centering
   \includegraphics[scale=0.96]{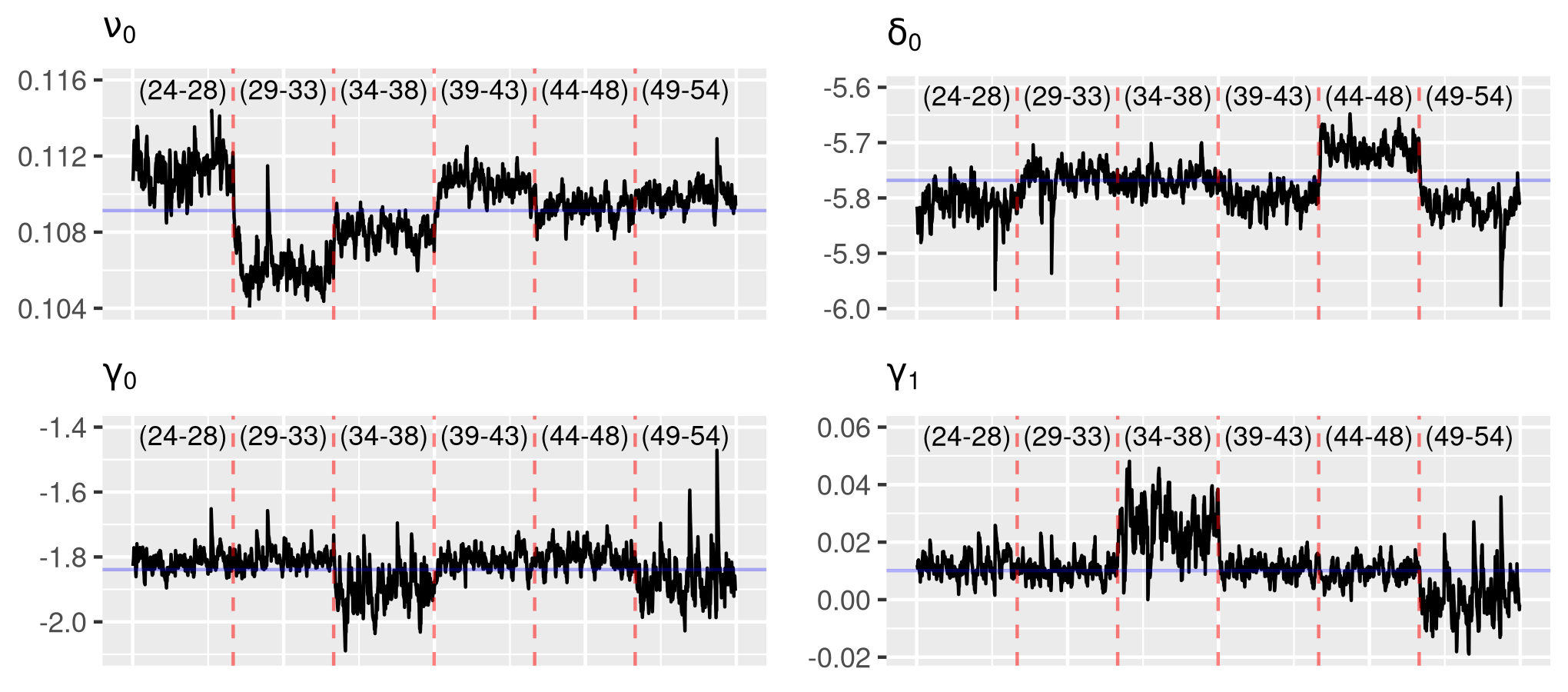}\\
   \textit{Legend: solid blue: full sample estimate; black line: \rnr\, draws excluding the age group indicated above in parenthesis; red dashed line: change in excluded age group}
\end{figure} 
Figure \ref{fig:moffitt}  presents results for estimation based on resampled data that exclude one age group at a time.  The parameter that is least sensitive to age-groups appears to be $\gamma_1$. The parameter $\nu_0$ tends to be lower when the age group (29-33) is excluded, while $\gamma_0$ is higher when the age group (44-48) is excluded. The parameter most sensitive to age is $\gamma_0$, which is evidently smaller in absolute magnitude when the younger age groups are excluded. For example, it is -1.2 when the youngest age group is dropped, but is around -2.0 when the oldest age group is excluded.

\paragraph{Application 3: Demand for Cereal}

We evaluate the \rnr\, algorithm on the BLP model of \citet{Berry1995} using the cereal data generated in \citet{nevo2000}. The sample consists of market shares for $24$ cereal products across $94$ markets for a total of $2256$ market/product observations. This example is of interest because bootstrapping the BLP model is demanding. We use the \textsc{BLPestimatoR} R package which builds on C++ functions to evaluate the GMM objective and analytical gradient. The data consists of market shares for 24 products in 94 markets. In the BLP framework, parameters on terms that enter linearly can projected out by 2SLS. Of the remaining parameters that need to be estimated,  we drop some interaction terms from the original paper that may be difficult to identify. This allows us to focus on coefficients that enter the moment conditions non-linearly since the BLP procedure requires a fixed-point inversion for these, making the moment conditions costly to evaluate. The parameter dimension including market fixed effects is $d=33$. To control for possible correlations in the unobservables at the market level, we compute cluster-robust standard errors. See Appendix \ref{apx:additional} for details

The data consists of market shares $s_{gj}$ in market $g \in \{1,\dots,94\}$ for product $j \in \{1,\dots,24\}$ with characteristic matrix $X_{gj}$. To resample at the market level, for each $b=1,\dots,B$ we draw markets $g^{(b)}_{1},\dots,g_{94}^{(b)}$ from $\{1,\dots,94\}$ with replacement and take the associated shares and characteristics $\{ s_{g^{(b)}j},X_{g^{(b)}j}\}_{j=1,\dots,24}$ as observations within each market. 
Since the number of clusers is relatively small, we only consider $m=n$. We set $\gamma = 0.2$ and a burn-in period of $10$ draws. Additional values of $\gamma$ are considered in Table \ref{tab:blp_gammas}.  
Both \rnr\, and r\textsc{qn} deliver  estimates  similar to those obtained from classical optimization and the standard errors are similar to the bootstrap ones.
However, there is a significant difference. To generate $B=1000$ draws the standard bootstrap requires 4h45m while the \rnr\, runs in 1h04m which is almost 5x faster. The  r\textsc{qn} estimate only requires 13m, which is about 20x faster than the bootstrap. While \dmk\, is similar to \rnr, a preliminary estimate of $\theta$ is needed.\footnote{\ks \; is not reported here because it needs significant rewriting of the \textsc{BLPestimatoR} package. Furthermore, \textsc{ks} only consider just-identified GMM models as indicated in their footnote 8.} Estimation of the  AR(1) coefficient for the \rnr\, draws associated with the parameters reported in Table \ref{tab:blp} finds that the estimates range  from $0.78$ to $0.82$ with 95\% confidence levels that always include the value of $(1-\gamma)=0.8$ predicted by theory.

{For this example we also consider the CH quasi-Bayesian estimator implemented using a random walk Metropolis-Hastings MCMC algorithm. The prior for each of the $33$ parameters is a $\mathcal{N}(0,100)$ distribution, and the inverse of the clustered variance-covariance matrix of the moments evaluated at $\hat\theta_n$ is used as the optimal weighting matrix $W_n$. 
The Markov chain is initialized at $\theta_0=\hat\theta_n$, the proposal in the random walk step is scaled by $0.2H_n(\hat\theta_n)^{-1/2}/\sqrt{n}$ which yields an acceptance rate of $0.335$. Though we generate a large number of draws ($B=50000$), the  Markov chain is strongly persistent and the effective sample size as defined in \citet{gelman2013}  is typically less than $100$. The CH estimates are further away from $\hat\theta_n$ than the \rnr\; estimates, and the standard errors are also smaller than the $m$ out of $n$ bootstrap and the \rnr\; or the \dmk.


\begin{table}[ht] \caption{Demand for Cereal: Estimates and Standard Errors (Random Coefficients)} \label{tab:blp}
      \centering
      \begin{tabular}{cl|bbaa|bbbaa}
        \hline \hline 
        & & \multicolumn{4}{c|}{Estimates} & \multicolumn{5}{c}{Standard Errors} \\ \hline
        & & $\hat\theta_n$ & \textsc{ch} & \mc{1}{\rnr} & \multicolumn{1}{c|}{r\textsc{qn}} & \textsc{ch} & \textsc{boot} & \dmk &  \mc{1}{\rnr} & \mc{1}{r\textsc{qn}} \\ 
        \hline \parbox[t]{2mm}{\multirow{4}{*}{\rotatebox[origin=c]{90}{stdev}}}
        & const. & 0.284 & -0.016 & 0.263 & 0.277 &  0.130 & 0.129 & 0.127  & 0.123 & 0.105 \\ 
        & price & 2.032 & 2.364 & 2.188 & 1.917 & 0.738 & 1.198 & 1.026  & 0.975 & 0.880 \\ 
        & sugar & -0.008 & -0.013 & -0.006 & -0.006 & 0.011 & 0.017 & 0.012 & 0.012 & 0.010 \\ 
        & mushy & -0.077 & -0.248 & -0.055 & -0.042 & 0.132 & 0.177 & 0.168 & 0.166 & 0.154 \\ \hline \parbox[t]{2mm}{\multirow{4}{*}{\rotatebox[origin=c]{90}{income}}}
        & const. & 3.581 & 4.414 & 3.464 & 3.702 & 0.453 & 0.666 & 0.738 & 0.714 & 0.636 \\ 
        & price & 0.467 & -3.255 & 1.335 & -0.295 & 2.449 & 3.829 & 4.275 & 4.040 & 3.603 \\ 
        & sugar & -0.172 & -0.195 & -0.171 & -0.174 & 0.021 & 0.028 & 0.028 & 0.027 & 0.025 \\ 
        & mushy & 0.690 & 0.888 & 0.647 & 0.702 & 0.203 & 0.345 & 0.346 & 0.339 & 0.312 \\  \hline
        \multicolumn{2}{c|}{} &  \mc{1}{} & \mc{1}{} & \mc{1}{} & \multicolumn{1}{c|}{time} & 1h50m &  4h45m & 1h4m & 1h8m & 13m \\      
         \hline \hline
      \end{tabular}
\end{table}

\section{Implications for Simulation-Based Estimation} \label{sec:smd}

Simulation-based estimation is routinely used to analyze structural models associated with analytically intractable likelihoods. Estimators in this class include the simulated method of moments, indirect inference and efficient method of moments, and following \citet{jjng-15}, we will generically refer to them as simulated minimum distance (\textsc{smd}) estimators.  We now show that  \rnr\ will still provide valid inference. This is useful because  computing  standard errors for the \textsc{smd} estimates is not always straightforward.

The minimum-distance (\textsc{md}) estimator minimizes the distance between a sample auxiliary statistic $\hat\psi_n=\hat\psi_n(\theta^0)$ and it's expected value $\psi(\theta)$  and is defined as
\[ \hat\theta_{n,\MD}=\text{argmin}_\theta\| \bar g_n(\theta)\|^2_{W_n}, \quad \bar g_n(\theta) = \hat\psi_n-\psi(\theta) \quad \tag{\textsc{md}} \]
 where $W_n$ is a weighting matrix. In cases when  the binding function $\psi(\cdot)$ that maps  $\theta$ to the auxiliary statistic is tractable, Algorithm \ref{algo:free_lunch} provides a convenient way to compute standard errors for $\hat\theta_{n,\MD}$. When $\psi(\theta)$ is  not tractable,  \textsc{smd} simulates data $y_{i,s}(\theta)$  for given $\theta$ using iid errors $e_{i,s}$ and estimates   $\psi(\theta)$ by  $\hat \psi_{n,S}(\theta)=\frac{1}{S}\sum_{s=1}^S \psi_{n,s}(y_{n,s}(\theta))$. The  estimator is
  \[ \hat\theta_{n,\textsc{smd}}=\text{argmin}_\theta \|\bar g_{n,S}(\theta)\|^2_{W_n} \quad \bar g_{n,S}(\theta)=  \hat{\psi}_n - \hat\psi_{n,S}(\theta) \quad \tag{\textsc{smd}} \]

To motivate our simulation based \rnr, consider  the exactly identified case when it holds that $\hat\psi_n-\psi(\hat\theta_{n,\MD})=0$.  Note that while  the vector of auxiliary statistics
  $\hat\psi_m^{(b)}$ computed using resampled data satisfies
  $ \mathbb{E}^\star ( \hat\psi_m^{(b)} ) = \hat\psi_n$, the statistics
 $\hat\psi_{m,S}^{(b)}(\theta)$ computed by \SMD\;  satisfies
  $ \mathbb{E}^\star [ \hat\psi_{m,S}^{(b)}(\theta)] = \psi(\theta)$, for all $\theta$.
 The two results together imply that 
  $ \mathbb{E}^\star [ \hat\psi_m^{(b)} - \hat\psi_{m,S}^{(b)}(\theta) ]$ when evaluated at  $\theta=\hat\theta_{n,\MD}$  is $\hat\psi_n-\psi(\hat\theta_{n,\MD})$ which takes the value of  zero as in  \MD\;  estimation. This suggests a simulation based  resampled objective function defined as: 
\[   Q_{m,S}^{(b)}(\theta)=\| \hat\psi_m^{(b)} -\hat \psi^{(b)}_{m,S}(\theta) \|^2_{W_n}, \quad \bar g_{m,S}(\theta)=\hat\psi_{m}^{(b)}-\hat\psi_{m,S}^{(b)}(\theta)\quad\tag{r\textsc{nr,s}}\]
will have the same minimizer as the infeasible \MD, at least to a first-order.
  Let the  draws be   generated according $\theta_{b+1,S}=\theta_{b,S} - \gamma P_{b+1,S} G_{m,S}^{(b)}(\thetab)$ with  gradient 
\begin{equation}
\label{eq:adjgrad1}
G_{m,S}^{(b)}(\theta_{b,S}) = -2 \partial_{\theta^\prime} \hat\psi_{m,S}^{(b)}(\theta_{b,S}) W_n (\hat\psi_m^{(b)}-\hat\psi_{m,S}^{(b)}(\theta)).
\end{equation}
By Theorem \ref{th:average_estimation}, the mean $\bar \theta_{\rnr,S}$ is consistent for $\hat\theta_{n,\MD}$.  By implication, $\theta_{\FL,S}$ will also be more efficient than $\hat\theta_{n,\SMD}$, which we will verify  in simulations below. To analyze  $\bar\theta_{\FL,S}$, we need the following:

\begin{assumptionbis}{ass:resampled} Suppose there exists finite constants $C_7,C_8$ such that for any $S \geq 1$
      \begin{itemize}
            \item[a.] $\left[ \mathbb{E}^\star \left(  \| \hat \psi_{m}^{(b)} - \hat \psi_{n} \|_2^4 \right) \right]^{1/4} \leq \frac{C_7}{\sqrt{m}}$; $\left[ \mathbb{E}^\star \left(  \sup_{\theta \in \Theta}\| \hat\psi_{m,S}^{(b)}(\theta) - \psi(\theta) \|_2^4 \right) \right]^{1/4} \leq \frac{C_7}{\sqrt{mS}}$
            \item[b.] $\left[ \mathbb{E}^\star \left(  \| \partial_\theta \hat\psi_{m,S}^{(b)}(\hat\theta_n) - \partial_\theta \psi(\hat\theta_n) \|_2^4  \right) \right]^{1/4} \leq \frac{C_8}{\sqrt{mS}}$.
      \end{itemize}
\end{assumptionbis}
Assumption \ref{ass:resampled}.iii' implies  Assumption \ref{ass:resampled}.iii where $G_n(\theta)$ is the gradient of \MD\; by taking the difference $G_{m,S}^{(b)}(\theta)-G_n(\theta)$ and using the Cauchy-Schwarz inequality. 




It remains to construct the variance of $\bar\theta_{\rnr,S}$.  The foregoing analysis would suggest that valid inference would follow after the variance adjustment defined in Algorithm \ref{algo:free_lunch}. However,  this is not the case.
Intuitively, the estimator $\overline{\theta}_{\rnr,S}$ is consistent for $\hat\theta_{n,\MD}$ whose variance $\mathbb V^0$  does not involve simulation noise. But  the quantity $ V_{\rnr}$ defined in Algorithm \ref{algo:free_lunch} presumes the presence of  simulation noise in the estimate $\overline{\theta}_{\rnr,S}$ and will  give  standard errors that will, in general, be too large. The are many ways to overcome this problem, and most  involve running a second chain of draws in parallel with the one used to compute the estimator. For example, taking the difference of two chains with the same simulated samples would work as the simulation noise will offset. 

Our preferred approach is to use a second chain that directly estimates the variance of $\bar\theta_{\rnr,S}$. As in the first chain, this second chain is generated as $\theta^2_{b+1,S}=(1-\gamma)\theta^2_{b,S}-\gamma P_{b+1,S}  \tilde{G}_{m,S}^{(b)}(\theta_{b,S}^2)$ as defined in (\ref{eq:lin}),  but the gradient is
\begin{equation}
\label{eq:adjgrad2}
\tilde{G}^{(b)}_{m,S}(\theta^1_{b,S})=-2\partial_{\theta^\prime} \hat\psi_{m,S}^{(b)}( \theta^1_{b,S} ) W_n ( \hat\psi_m^{(b)}-\hat\psi_n ).
\end{equation}
Compared to the first chain defined by (\ref{eq:adjgrad1}), the second  chain  replaces  the simulated auxiliary statistics $\hat \psi_{m,S}^{(b)}(\hat\theta_n)$ by the sample estimates $\hat\psi_{n}$ which is already computed. As  all other quantities involved in computing (\ref{eq:adjgrad2}) are taken from (\ref{eq:adjgrad1}),  the computation overhead of generating $\theta^2_{b,S}$ is thus negligible.

\begin{proposition} \label{cor:smd_rnr} Suppose that the Assumptions for $Q_n$ and $Q^{(b)}_{m,S}$ in Theorems \ref{th:average_estimation} and \ref{th:free_lunch} hold,  with  Assumption \ref{ass:resampled}.iii  replaced by \ref{ass:resampled}.iii'. Let  $\hat \theta_{n,\MD} $ be the infeasible minimum-distance estimator. Let   $\{\theta^1_{b,S}\}$  be a chain generated with $G^{(b)}_{m,S}$  defined as in (\ref{eq:adjgrad1}), and $\{\theta^2_{b,S}\}$ be  generated using $\tilde{G}_{m,S}^{(b)}$  defined as in (\ref{eq:adjgrad2}).  Let $\overline{\theta}_{\rnr,S} = \frac{1}{B} \sum_{b=1}^B \theta_{b,S}^1$,  $P_{b+1,S} = [H_{m,S}^{(b+1)}(\theta^1_{b,S})]^{-1}$, and define
      $\mathbb V_{\rnr,S} = \frac{m}{\phi(\gamma)}\var^\star(\theta_{b,S}^2)$. Then for any $S\geq 1$ fixed,
\begin{itemize}
      \item[i.]       
      $\sqrt{n}\left( \overline{\theta}_{\rnr,S} - \theta^0 \right) = \sqrt{n}\left( \hat\theta_{n,\MD} - \theta^0 \right) + o_{p^\star}(1)$.
      \item[ii.]  
      As $m,b \to +\infty$ with $\log(m)/b \to 0$:
           \begin{align*} 
                 \mathbb V_{\rnr,S}^{-1/2} \sqrt{n}\left(\overline \theta_{\rnr,S} - \theta^0 \right) \overset{d^\star}{\to} \mathcal{N}\left( 0, I_d \right),
           \end{align*}  
\end{itemize}
\end{proposition}
  
\citet{jjng-15,jjng-14} shows that a weighted average of \textsc{smd} estimates with independent simulation draws constitutes a posterior mean which is asymptotically equivalent to the infeasible \textsc{md} estimator. This requires solving the optimization problem as many times (ie.  $S>1$). Part i. of the proposition shows that this type of statistical efficiency can be achieved by \rnr\; in a single run, ie($S=1$). Resampling by \rnr\; involves taking draws from the joint distribution $F_n \times F_{\text{shocks}}$ to produce $\hat \psi^{(b)}_{m,S}(\theta)$, which is an estimate of population mapping $\psi(\theta)$.  In practice, the simulation and resampling noise in $\hat \psi^{(b)}_{m,S}(\theta)$ and $\hat\psi_m^{(b)}$ are averaged out so that the variance of $\overline{\theta}_{\rnr,S}$ does not depend on $S$ asymptotically. This contrasts with the \textsc{smd} estimator $\hat \theta_{n,\SMD}$  which has vanishing simulation noise only when $S\rightarrow\infty$ as $n \to \infty$.  

Part ii of the Proposition involves a second sequence $\theta^2_{b,S}$ which, as noted above, is used to  compute the variance of the estimator.  To understand its underpinnings, recall that the  sandwich variance  for $\hat\theta_{n,\MD}$ has a meat component that is  the variance of the score $-2\partial_{\theta^\prime} \psi ( \hat \theta_{n,\MD} ) W_n ( \hat\psi_n-\psi (\hat\theta_{n,\MD}) )$. If $\psi$ were tractable, a bootstrap draw of this score would be  $-2\partial_{\theta^\prime} \psi ( \hat \theta_{n,\MD} ) W_n ( \hat\psi_m^{(b)}-\psi (\hat\theta_{n,\MD}) )$. But this is approximately $-2\partial_{\theta^\prime} \hat\psi_{m,S}^{(b)}( \theta_{b,S}^1 ) W_n ( \hat\psi_m^{(b)}-\hat\psi_n )$ which is precisely  the gradient (\ref{eq:adjgrad2}) used to generate $\theta_{b,S}^2$. Hence it  provides a correct  approximation of the variance of the scores. Though two chains are needed in the case of simulation estimation, it  only needs $S=1$. These arguments are further illustrated using a simple example in Appendix \ref{apx:smd}.

\paragraph{Example 3: Dynamic Panel} 
Consider the dynamic panel regression:
\[ y_{it} = \alpha_i + \rho y_{it-1} + x_{it}^\prime\beta +\sigma_e e_{it},  \]
with $\rho = 0.6,\beta=1,\sigma_e=1$, $x_{it} \sim \mathcal{N}(0,1), e \sim \mathcal{N}(0,1)$, $n=1000$ and $T=5$. 
Let $A = I_T - 1_T 1_T^\prime /T$, a matrix which computes the time de-meaned $y_{it} - \bar y_i$. The Least-Squares Dummy Variable (LSDV) estimator is obtained by regressing $A y_{2:T}$ on $A y_{1:T-1}$ and $A x_{2:T}$. The estimator is inconsistent for fixed $T$ as $n \to \infty$.

The LSDV estimator is inconsistent when $n\to\infty$ and $T$ is fixed. \citet{Gourieroux2010} shows that indirect inference, which has an automatic bias correction property, is consistent for fixed $T$. The idea is to match the sample LSDV estimator $\hat\psi_n = \hat\theta_{n,LSDV}$ with a simulated $\hat\psi_{n,S}(\theta) = \hat \theta^{sim}_{n,LSDV}(\theta)$ using $S\geq 1$ simulated samples. 



To generate  \rnr\, draws, we resample  $(y_{i1},\dots,y_{iT},x_{i1},\dots,x_{iT})_{i=1,\dots,m}$ with replacement over $i$ for given $m$  and compute $\hat\psi_m^{(b)}=\hat \theta_{m,LSDV}^{(b)}$, our resampled moments. Using the new simulation draws $e^{(b+1)}_{it}$ at each $b$, we simulate $S \geq 1$ panels: 
$y^{(b+1)}_{it,s} = \rho y_{it-1,s} + x_{it,s}^{ (b+1) \prime}\beta +\sigma_e e^{(b+1)}_{it,s}$, for $t=1,\dots,T$ and $i=1,\dots,m$ and compute the simulated moments $\hat\psi_{m,S}^{(b)}=\tilde \theta_{m,LSDV}^{(b)}(\thetab)$.  An addional moment is needed to estimate $\sigma_e$;  we use the standard deviation of the OLS residuals in the LSVD regression. The gradient and Hessian are computed using finite differences. We illustrate with  $m=n,100,50$ for $n=1000$.

\begin{table}[ht]
      \centering \caption{Dynamic Panel: Estimates of $\rho$ and Standard Errors} \label{tab:panel}
{ 
\begin{tabular}{ll|baaa|bbbaaa}
      \hline \hline
     & &  \multicolumn{4}{c|}{Estimates}  &  \multicolumn{6}{c}{Standard Errors} \\ 
   $S$ &  $m$  & \mc{1}{\textsc{ind}} & \mc{1}{\rnr$_{0.3}$} & \mc{1}{\rnr$_{0.1}$} & \multicolumn{1}{c|}{\rnr$_{0.01}$} & \textsc{ase} & \textsc{boot} & \dmk & \mc{1}{\rnr$_{0.3}$} & \mc{1}{\rnr$_{0.1}$} & \mc{1}{\rnr$_{0.01}$}\\
        \hline
        & $200$ & 0.619 & 0.589 & 0.592 & 0.590 & 0.045 & 0.049 & 0.050 & 0.034 & 0.034 & 0.025 \\ 
       1 & $100$ & - & 0.589 & 0.588 & 0.586 & - & 0.048 & - & 0.036 & 0.039 & 0.037 \\ 
        & $50$ & - & 0.580 & 0.588 & 0.581 & - & 0.050 & - & 0.037 & 0.037 & 0.024 \\  \hline
        & $200$ & 0.584 & 0.591 & 0.589 & 0.589 & 0.035 & 0.036 & 0.036 & 0.035 & 0.036 & 0.032 \\ 
       10 & $100$ & - & 0.589 & 0.591 & 0.587 & - & 0.034 & - & 0.034 & 0.037 & 0.030 \\ 
        & $50$ & - & 0.586 & 0.589 & 0.587 & - & 0.035 & - & 0.038 & 0.031 & 0.032 \\ 
         \hline \hline
      \end{tabular}
      }\\
      \textit{Remark: Results reported for one simulated sample of size $n=200, T=5$.}
\end{table}




The LSDV estimate is 0.329 which is significantly downward biased. However,  the indirect inference (\textsc{ind}) estimator corrects the bias as shown in  \citet{Gourieroux2010}. The estimate of $0.619$ in Table \ref{tab:panel} for $S=1$ bears this out. The \rnr\, estimates are closer to $\theta^0$ than \textsc{ind} for $m=n$, $100$ and is similar for $m=50$. The \textsc{ind} estimates with $S=10$ are very close to the \rnr\; estimates obtained over all $\gamma, m$ and $S$ including $S=1$. This implies that \rnr\; achieves the efficiency of \textsc{ind} with large $S$ using just $S=1$. The standard errors are smaller than other methods except for $S=10$. Results for $S=2,5$ are reported in Table \ref{tab:panel_apx}.

We close the analysis with two remarks about the examples.  As noted earlier, an ill-conditioned Hessian can render slow convergence of gradient-based optimizers. The values of $\frac{\lambda_\min(H_n)}{\lambda_\max(H_n)}$  evaluated at $\theta=\hat\theta_n$, are $ 10^{-7},8 \cdot 10^{-4}$ and $7 \cdot 10^{-6}$ for the probit, earnings dynamics, and BLP  examples, respectively. Classical \gd\ should be slow in converging in these cases, and the applications bear this out.
Second, to reinforce the main result  that Algorithm \ref{algo:free_lunch} provides valid inference, we evaluate  the coverage of \rnr\;  in all of the simulated examples considered.  As seen from Table \ref{tab:coverage}, \rnr\; delivers a  5\% size in almost all cases. Details are given in  Appendix \ref{apx:additional} of the online supplement.

\begin{table}[H] \caption{Size of Confidence Intervals Across Methods and Examples} 
      \label{tab:coverage}
      \centering
      { 
      \begin{tabular}{cl|c|bbba|ba}
            \hline \hline
            \multicolumn{2}{c|}{} & \textsc{ase} & \textsc{boot} & \dmk & \ks & \multicolumn{1}{c|}{\rnr} &  \textsc{boot} & \mc{1}{\rnr} \\ \hline
            \multirow{3}{*}{OLS} & & &  \multicolumn{4}{c|}{$m=n=200$} & \multicolumn{2}{c}{$m=50$}\\
            & $\beta_0$ & 0.044 & 0.043 & 0.043 & 0.041 & 0.049 & 0.040 & 0.047 \\ 
            & $\beta_1$ & 0.045 & 0.056 & 0.056 & 0.070 & 0.069 & 0.043 & 0.048  \\ \hline
            \multirow{3}{*}{MA(1)} & & &  \multicolumn{4}{c|}{$m=n=500$} & \multicolumn{2}{c}{$m=250$}\\
            & $\mu$ & 0.291 & 0.048 & 0.294 & - & 0.183 & 0.051 & 0.169  \\ 
            & $\psi$ & 0.066 & 0.047 & 0.067 & - & 0.064 & 0.035 & 0.044  \\ \hline
            & & &  \multicolumn{4}{c|}{$m=n=1000$} & \multicolumn{2}{c}{$m=100$}\\
            Dynamic & $\rho$ & 0.055 & 0.047 & 0.044 & - & 0.050 & 0.052 & 0.040 \\ 
            Panel & $\beta$ & 0.055 & 0.054 & 0.051 & - & 0.057 & 0.051 & 0.049  \\
            ($S=1$) & $\sigma$ & 0.060 & 0.053 & 0.046 & - & 0.057 & 0.052 & 0.059  \\
            \hline
            & & &  \multicolumn{4}{c|}{$m=n=1000$} & \multicolumn{2}{c}{$m=100$}\\
            Dynamic & $\rho$ & 0.051 & 0.054 & 0.055 & - & 0.053 & 0.059 & 0.053 \\ 
            Panel & $\beta$ & 0.040 & 0.046 & 0.046 & - & 0.049 & 0.045 & 0.049  \\
            ($S=2$) & $\sigma$ & 0.065 & 0.056 & 0.053 & - & 0.056 & 0.056 & 0.056  \\
            \hline
            & & &  \multicolumn{4}{c|}{$m=n=1000$} & \multicolumn{2}{c}{$m=100$}\\
            Dynamic & $\rho$ & 0.052 & 0.054 & 0.053 & - & 0.048 & 0.051 & 0.050 \\ 
            Panel & $\beta$ & 0.040 & 0.047 & 0.042 & - & 0.036 & 0.043 & 0.038 \\
            ($S=5$) & $\sigma$ & 0.065 & 0.058 & 0.056 & - & 0.064 & 0.061 & 0.061 \\
            \hline \hline
      \end{tabular}
      }\\
\textit{Results based on $1000$ replications with $B=1000, \gamma = 0.1$; \textsc{burn}=45.}
\end{table}

\section{Conclusion}

In this paper, we design  two algorithms to produce draws that, upon averaging, is asymptotically equivalent to the  full-sample estimate produced by a classical optimizer. By using the inverse Hessian as conditioning matrix, the draws of Algorithm 2 immediately provide valid  standard errors for inference, hence a  free lunch.   In problems that require  $S$  simulations to approximate the binding function, our algorithm achieves the level of efficiency of \textsc{smd}  with a large $S$, but  at the cost of $S=1$.  Numerical evaluations show that Algorithm 2 produces accurate estimates and standard errors but   runs  significantly faster than the conventional bootstrap and  most of the `short-cut' methods. 


\newpage
\baselineskip=12.0pt
\bibliographystyle{ecta}
\bibliography{refs}

\begin{appendices}
      \renewcommand\thetable{\thesection\arabic{table}}
      \renewcommand\thefigure{\thesection\arabic{figure}}
      \renewcommand{\theequation}{\thesection.\arabic{equation}}
      \renewcommand\thelemma{\thesection\arabic{lemma}}
      \renewcommand\thetheorem{\thesection\arabic{theorem}}
      \renewcommand\thedefinition{\thesection\arabic{definition}}
        \renewcommand\theassumption{\thesection\arabic{assumption}}
      \renewcommand\theproposition{\thesection\arabic{proposition}}
        \renewcommand\theremark{\thesection\arabic{remark}}
        \renewcommand\thecorollary{\thesection\arabic{corollary}}
\setcounter{equation}{0}
\clearpage \baselineskip=18.0pt
\setcounter{page}{1}
\renewcommand*{\thepage}{A-\arabic{page}}
\appendix
\section{} \label{apx:proofs}
\subsection{Derivations for the Least-Squares Example} \label{apx:OLS}
In this example, $y_n = X_n \hat \theta_n + \hat e_n$, $Q_n(\theta) = \frac{1}{2n} (y_n  -X_n\theta)^\prime(y_n  -X_n\theta)$,  $H_n=X_n'X_n/n, G_n=-X_n^\prime \hat e_n/n$, and $Q_m^{(b)}(\theta) = \frac{1}{2m} (y_m^{(b)}  -X_m^{(b)}\theta)^\prime(y_m^{(b)}  -X_m^{(b)}\theta)$, $H_b=H^{(b+1)}_m(\theta_b)=X^{(b+1)^\prime}_mX^{(b+1)}_m/m$. $G_b(\theta) = -X_m^{(b+1)^\prime}[y_m^{(b+1)} - X_m^{(b+1)}\theta]/m$. 
Let $\hat \theta_m^{(b+1)} = (X_m^{(b+1) \prime} X_m^{(b+1)})^{-1} X_m^{(b+1) \prime} y_{m}^{(b+1)}$ be the $m$ out of $n$ bootstrap estimate. Orthogonality of least squares  residuals 
will be used repeatedly.
\paragraph{Gradient Descent}
$ \theta_{k+1} = \theta_k - \gamma \left[ -X_n^\prime (y_n-X_n\theta_k)/n \right]$.
Subtract $\hat \theta_n$ on both sides and note that  $y_n = X_n \hat \theta_n + \hat e_n$ (full sample estimates), then:
\begin{align*}
 \theta_{k+1} - \hat \theta_n &= \theta_k - \hat \theta_n - \gamma \left[ -X_n^\prime (X_n\hat\theta_n + \hat e_n -X_n\theta_k)/n \right]\\
 &= \theta_k - \hat \theta_n - ( \gamma H_n) (\theta_k - \hat \theta_n)  + \gamma X_n^\prime \hat e_n/n= (I- \gamma H_n) (\thetab - \hat \theta_n) \quad \text{since } X_n^\prime\hat e_n=0.
\end{align*}

\paragraph{Newton-Raphson} 
$ \theta_{k+1} = \theta_k - \gamma \left[ H_n \right]^{-1} \left[ -X_n^\prime (y_n-X_n\theta_k)/n \right].$
Subtract $\hat \theta_n$ on both sides:
\begin{align*}
 \theta_{k+1} - \hat \theta_n &= \theta_k - \hat \theta_n - \gamma H_n^{-1}\left[ -[X_n^\prime X_n/n][\hat \theta_n-\theta_k] +  X_n^\prime\hat e_n/n  \right]
 = (1-\gamma)(\theta_k - \hat \theta_n) \quad \text{since } X_n^\prime \hat e_n=0.
\end{align*}

\paragraph{Stochastic Gradient Descent} 
$ \thetabf = \thetab - \gamma_b \left[ - X_m^{(b) \prime} (y_m^{(b)}-X_m^{(b)}\thetab )/m \right]$. Thus
\begin{align*}
      \thetabf - \hat \theta_n &= \thetab - \hat \theta_n - \gamma_b \left[ -X_m^{ (b+1) \prime} (y_m^{(b+1)}- X_m^{(b+1)}\hat\theta_n  - X_m^{(b+1)}[\thetab-\hat\theta_n])/m \right]\\
      &= (I- \gamma_b H_b) (\thetab - \hat \theta_n)  + \gamma_b X_m^{(b+1) \prime }(y_m^{(b+1)}- X_m^{(b+1)}\hat\theta_n )/m\\
      &= (I- \gamma_b H_b) (\thetab - \hat \theta_n)  - \gamma_b G_b(\hat\theta_n)\quad \text{since } X_m^{(b+1)^\prime} \hat e_m^{(b+1)}=0.
     \end{align*}

\paragraph{Resampled Gradient Descent} 
$ \thetabf = \thetab - \gamma \left[ - X_m^{(b+1) \prime} (y_m^{(b+1)}-X_m^{(b+1)}\thetab )/m \right]$.
Subtract $\hat \theta_n$ on both sides and note that  $y_m^{(b+1)} = X_m^{(b+1)} \hat \theta_m^{(b+1)} + \hat e_m^{(b+1)}$ (bootstrap estimates). Then
\begin{align*}
      \thetabf - \hat \theta_n &= \thetab - \hat \theta_n - \gamma \left[ -X_m^{ (b+1) \prime} (X_m^{(b+1) }[\hat\theta_m^{(b+1)}-\hat\theta_n]  + \hat e_m^{(b+1)} -X_m^{(b+1) }[\thetab-\hat\theta_n])/m \right]\\
      &= \thetab - \hat \theta_n - ( \gamma H_b) (\thetab - \hat \theta_n)  + \gamma H_b (\hat\theta_m^{(b)} -\hat\theta_n)\\ &= (I- \gamma H_b) (\thetab - \hat \theta_n) + \gamma H_b (\hat\theta_m^{(b+1)} -\hat\theta_n) \quad \text{since } X_m^{(b+1)^\prime} \hat e_m^{(b+1)}=0.
     \end{align*}
\paragraph{Resampled Newton-Raphson} 
$ \thetabf = \thetab - \gamma [H_b]^{-1} \left[ - X_m^{(b+1) \prime} (y_m^{(b+1)}-X_m^{(b+1)}\thetab )/m \right]$. Then
     \begin{align*}
           \thetabf - \hat \theta_n &= \thetab - \hat \theta_n - \gamma [H_b]^{-1} \left[ -X_m^{ (b+1) \prime} (X_m^{(b+1) }[\hat\theta_m^{(b+1)}-\hat\theta_n]  + \hat e_m^{(b+1)} -X_m^{(b+1) }[\thetab-\hat\theta_n])/m \right]\\
           &= (1- \gamma) (\thetab - \hat \theta_n) + \gamma (\hat\theta_m^{(b+1)} -\hat\theta_n) \quad \text{since } X_m^{(b+1) \prime} \hat e_m^{(b+1)}=0.
          \end{align*}

\subsection{Proof of Lemma \ref{lem:coupling}:}
Note first that by construction,
\begin{align}
      \gamma &\left( \pb G_m^{(b+1)}(\thetab) - \overline{P}_m G_m^{(b+1)}(\hat\theta_n) \right)  =\gamma \overline{P}_m H_n(\hat\theta_n)[\thetab - \hat\theta_n] \nonumber \\ &\quad + \gamma\overline{P}_m \left( G_m^{(b+1)}(\thetab)-G_m^{(b+1)}(\hat\theta_n)- H_n(\hat\theta_n)[\thetab - \hat\theta_n]\right) \label{eq:d3}
      \\ &\quad +\gamma \left( \pb - \overline{P}_m\right) \left( G_m^{(b+1)}(\thetab)-G_m^{(b+1)}(\hat\theta_n) \right). \label{eq:d4}
\end{align}
From the definition of $\thetab$ and $\thetab^\star$, the difference can be expressed as:
\begin{align*}
      \thetabf-\thetabf^\star &= \left(\thetab - \gamma \pb G_m^{(b+1)}(\thetab) \right) - \left( \hat\theta_n + \Psi(\hat\theta_n)(\thetab^\star - \hat\theta_n) - \gamma \overline{P}_m G_m^{(b+1)}(\hat\theta_n) \right) \nonumber\\
      &= \Psi(\hat\theta_n)(\thetab-\thetab^\star)  + (I_d-\Psi(\hat\theta_n))(\thetab-\hat\theta_n)  -\gamma \left( \pb G_m^{(b+1)}(\thetab) - \overline{P}_m G_m^{(b+1)}(\hat\theta_n) \right)\\
&= \Psi(\hat\theta_n)(\thetab-\thetab^\star)+\gamma\overline{P}_m H_n(\hat\theta_n)[\thetab-\hat\theta_n]-
\gamma \left( \pb G_m^{(b+1)}(\thetab) - \overline{P}_m G_m^{(b+1)}(\hat\theta_n) \right)\\
&=\Psi(\hat\theta_n)(\thetab-\thetab^\star)- (\ref{eq:d3})- (\ref{eq:d4})
\end{align*}
where the third equality follows from the fact that $I_d-\Psi(\hat\theta_n) = \gamma\overline{P}_m H_n(\hat\theta_n)$.  By Assumption \ref{ass:resampled} i. and vi. as well as Lemma \ref{lem:cv_stochastic}, 
\begin{align*}
\mathbb{E}^\star(\| (\ref{eq:d3}) \|_2) &\leq \gamma \overline{\lambda}_P C_2 \mathbb{E}^\star(\|\thetab -\hat\theta_n\|_2^2)\\ &\leq 3\gamma \overline{\lambda}_P C_2 \left(  (1-\overline{\gamma})^{2b+2}d_{0,n}^2 + \frac{C_5^2}{\overline{\gamma}^2m}\right).
\end{align*}
By Assumptions \ref{ass:resampled} ii., \ref{ass:K} ii., Lemma \ref{lem:cv_stochastic},  mean-value theorem, and  Cauchy-Schwarz inequality, 
\begin{align*}
      \mathbb{E}^\star \left( \| (\ref{eq:d4}) \|_2\right)& \leq \gamma\left[ \mathbb{E}^\star \left( \|\pb-\overline{P_m}\|_2^2 \right) \right]^{1/2} \left[ \mathbb{E}^\star \left( \| H_m^{(b+1)}(\tilde{\theta}_b)(\thetab - \hat\theta_n)\|_2^2 \right) \right]^{1/2}\\
      &\leq \gamma\overline{\lambda}_HC_6 \left( \overline{\rho}^b d_{0,n} + \frac{1}{\sqrt{m}} \right)  \left( (1-\overline{\gamma})^{b+1} d_{0,n} + \frac{C_5}{\overline{\gamma}\sqrt{m}} \right),
\end{align*}
where $\tilde \theta_b$ is some intermediate value between $\thetab$ and $\hat \theta_n$, and  an upper bound defined in terms of $\bar\rho$ to simplify notation.

The two bounds leads to the following recursion on the coupling distance:
\begin{align*}
      \mathbb{E}^\star \left( \|\thetabf - \thetabf^\star \|_2\right) &\leq \overline{\rho} \mathbb{E}^\star \left( \|\thetab - \thetab^\star \|_2\right) + \mathbb{E}^\star (\| (\ref{eq:d3}) \|_2) + \mathbb{E}^\star (\| (\ref{eq:d4}) \|_2)\\
      &\leq \overline{\rho} \mathbb{E}^\star \left( \|\thetab - \thetab^\star \|_2\right) + C_6^+(\overline{\rho}^b [d_{0,n}+d_{0,n}^2] + \frac{1}{m})\\
      &\leq \frac{C_6^+}{1-\overline{\rho}} \left( \overline{\rho}^b [d_{0,n}+d_{0,n}^2] + \frac{1}{m}\right),
\end{align*}
where $C_6^+$ is a constant which depends on the terms used to bound (\ref{eq:d3}) and (\ref{eq:d4}).   Recall that $\theta_0 = \theta_0^\star$ so that the coupling distance is zero for $b=0$. Putting  $C_7 = C_6^+/(1-\overline{\rho})$ proves the desired result.\qed\\

\subsection{Proof of Theorem \ref{th:average_estimation}}
To bound $\mathbb{E}^\star\left( \| \overline{\theta}_{\FL}^\star - \hat\theta_n \|_2 \right)$, we use the recursive representation of (\ref{eq:lin}) and take the average:
\begin{align*}
      \overline{\theta}_{\FL}^\star- \hat\theta_n &=  \frac{1}{B} \sum_{b=1}^B\Psi(\hat\theta_n)^{b}(\theta_0-\hat\theta_n) -\gamma \frac{1}{B} \sum_{b=1}^B \sum_{j=0}^{b-1} \Psi(\hat\theta_n)^j\overline{P}_m \mathbb{E}^\star \left(G_m^{(b-j)}(\hat\theta_n)\right) \\ &-\gamma \frac{1}{B} \sum_{b=1}^B \sum_{j=0}^{b-1} \Psi(\hat\theta_n)^j\overline{P}_m \underbrace{\left[ G_m^{(b-j)}(\hat\theta_n)  - \mathbb{E}^\star \left(G_m^{(b-j)}(\hat\theta_n)\right) \right]}_{\Delta_m^{(b-j)}(\hat\theta_n)}.
\end{align*}
Assumption \ref{ass:K} i. implies that $\|\Psi(\hat\theta_n)^{b}(\theta_0-\hat\theta_n)\|_2 \leq \overline{\rho}^{b}\|\theta_0-\hat\theta_n\|_2$, so the first term is less than $\frac{d_{0,n}}{(1-\overline{\rho})B}$ in expectation. Consider now the second term. By Assumption \ref{ass:resampled} iv, $\|\Psi(\hat\theta_n)^j\overline{P}_m \mathbb{E}^\star \left(G_m^{(b-j)}(\hat\theta_n)\right)\|_2 \leq \overline{\rho}^j \overline{\lambda}_P \frac{C_3^\prime}{\sqrt{m}}$ so  the second term is less than $\frac{\overline{\lambda}_P C_3^\prime }{(1-\overline{\rho})\sqrt{m}}$. For the third term and with $\Delta_m^{(b-j)}(\hat\theta_n)$ defined above, we have by conditional independence,
{\small
\begin{align*}
\bigg[\mathbb{E}^\star \left( \| \frac{1}{B} \sum_{b=1}^B \sum_{j=0}^{b-1} \Psi(\hat\theta_n)^j \overline{P}_m\Delta_m^{(b-j)}(\hat\theta_n) \|_2^2 \right)\bigg]^{1/2}
&= \bigg[\mathbb{E}^\star \left( \| \frac{1}{B} \sum_{b=1}^B \sum_{j=0}^{B-b+1} \Psi(\hat\theta_n)^j \overline{P}_m\Delta_m^{(b)}(\hat\theta_n) \|_2^2 \right)\bigg]^{1/2}\\
&= \frac{1}{\sqrt{mB}}
\bigg[ \frac{1}{B}  \sum_{b=1}^{B} \mathbb{E}^\star \left(  \| \sum_{j=0}^{B-b+1} \Psi(\hat\theta_n)^j \sqrt{m}\overline{P}_m \Delta_m^{(b)}(\hat\theta_n) \|_2^2 \right)\bigg]^{1/2}\\
&\le \frac{\overline{\lambda}_P}{(1-\overline{\rho})\sqrt{mB}}\bigg[ \left( \sup_{1 \leq b \leq B} \mathbb{E}^\star  \|  \sqrt{m}\Delta_m^{(b)}(\hat\theta_n)  \|_2^2 \right)\bigg]^{1/2}\\
&\leq  \frac{\gamma \overline{\lambda}_P [C_3+C_3^\prime]}{(1-\overline{\rho})\sqrt{mB}}
\end{align*}
}
where the first inequality follows from the average being less than the $\sup$, combined with $\| \Psi(\hat\theta_n)^j \overline{P}_m\Delta_m^{(b)}(\hat\theta_n) \| \leq \overline{\rho}^j \overline{\lambda}_P \|\Delta_m^{(b)}(\hat\theta_n) \|$ which is summable over $j \geq 0$. The last inequality uses Assumption \ref{ass:resampled} iii-iv. 
Recall that Lemma \ref{lem:coupling} implies (\ref{eq:avg_coupling}) which states that
$ \mathbb E^\star \bigg(\|\overline{\theta}_{\FL}-\bar \theta^\star_{\FL}\|_2\bigg)\le \frac{C_7 }{1-\overline{\rho}}\bigg(\frac{1}{m}+\frac{d_{0,n}+d_{0,n}^2}{B} \bigg)$. Now putting everything together, we have:
\begin{align*} \mathbb{E}^\star\left( \| \overline{\theta}_{\FL} - \hat\theta_n \|_2 \right) &\leq \mathbb E^*\bigg(\|\bar\theta_{\FL}-\overline\theta_{\FL}^*\|_2\bigg)+\mathbb E^*\bigg(\overline\theta_{\FL}^*-\hat\theta_n\|_2\bigg)\\
& \leq C_8 \left( \frac{1}{m} + \frac{d_{0,n}+d_{0,n}^2}{B} + \frac{1}{\sqrt{mB}} \right), 
\end{align*}
which is a $o(\frac{1}{\sqrt{n}})$ when $\frac{\sqrt{n}}{\min(m,B)} \to 0$ and $d_{0,n} = O(1)$.\qed\\

\subsection{Proof of Theorem \ref{th:free_lunch}}
The property that $\overline{P}_m = [H_n(\hat\theta_n)]^{-1}$ when  $\pb = [H_m^{(b+1)}(\thetab)]^{-1}$ is crucial for what is to follow, and it is useful to understand why.  Under Assumption \ref{ass:resampled} vi., 
\[ \left[\mathbb{E}^\star \left( \|I_d-\pb H_n(\hat\theta_n)\|_2^2 \right) \right]^{1/2} \leq \frac{1}{\underline{\lambda}_P} \left[\mathbb{E}^\star \left( \|\pb^{-1} - H_n(\hat\theta_n) \|_2^2 \right) \right]^{1/2}. \]
Given that $\pb = [H_m^{(b+1)}(\thetab)]^{-1}$, an application of the triangular inequality,  Assumption \ref{ass:Qn} ii. and \ref{ass:resampled} v. together with Lemma \ref{lem:cv_stochastic} give
\begin{align*}
      \left[\mathbb{E}^\star \left( \|\pb^{-1} - H_n(\hat\theta_n) \|_2^2 \right) \right]^{1/2} &= \left[\mathbb{E}^\star \left( \|H_m^{(b)}(\thetab) - H_n(\hat\theta_n) \|_2^2 \right) \right]^{1/2}\\&\leq  \left[\mathbb{E}^\star \left( \|H_n(\thetab) - H_n(\hat\theta_n) \|_2^2 \right) \right]^{1/2} + \left[\mathbb{E}^\star \left( \|H_m^{(b+1)}(\thetab) - H_n(\thetab) \|_2^2 \right) \right]^{1/2}\\
      &\leq C_1 \left[\mathbb{E}^\star \left( \| \thetab - \hat\theta_n\|_2^2 \right) \right]^{1/2} + \left[\mathbb{E}^\star \left( \sup_{\theta \in \Theta} \|H_m^{(b+1)}(\theta) - H_n(\theta) \|_2^2 \right) \right]^{1/2}\\
      &\leq (1-\overline{\gamma})^{b}C_1d_{0,n} + \left( \frac{C_1 C_5}{\overline{\gamma} } + C_4 \right)\frac{1}{\sqrt{m}}.
\end{align*}
This implies that Assumption \ref{ass:K} ii. holds with $C_6 = \max(C_1,\frac{C_1 C_5}{\overline{\gamma} } + C_4)$ and $\overline{P}_m = [H_n(\hat\theta_n)]^{-1}$.  Assumption \ref{ass:K} i. automatically holds since we now have $\Psi(\hat\theta_n)=(1-\gamma)I_d$ which has all its eigenvalues in $[0,1)$ for any $\gamma \in (0,1]$.

To prove Theorem \ref{th:free_lunch},  we first substitute $\thetab$ for the linear process $\thetab^\star$ using:
\begin{align*}
      \frac{\sqrt{m}}{\sqrt{\phi(\gamma)}}(\mathbb{V}^m)^{-1/2}(\thetab-\hat\theta_n) = \frac{\sqrt{m}}{\sqrt{\phi(\gamma)}}(\mathbb{V}^m)^{-1/2}(\thetab-\thetab^\star) + \frac{\sqrt{m}}{\sqrt{\phi(\gamma)}}(\mathbb{V}^m)^{-1/2}(\thetab^\star-\hat\theta_n).
\end{align*}
By Lemma \ref{lem:coupling}, $\frac{\sqrt{m}}{\sqrt{\phi(\gamma)}}(\mathbb{V}^m)^{-1/2}(\thetab-\thetab^\star) = o_{p^\star}(1)$ when $\log(m)/b \to 0$ since it implies $\sqrt{m}\overline{\gamma}^b = \exp( b[\frac{\log(m)}{2b}+\log(\overline{\gamma})]) \to 0$.

For \rnr\, we have $\overline{P}_m = H_n(\hat\theta_n)^{-1}$ so that $\Psi(\hat\theta_n) = (1-\gamma)I_d$. Using the recursion (\ref{eq:lin}), we have:
\begin{align*}
      \frac{\sqrt{m}}{\sqrt{\phi(\gamma)}}(\mathbb{V}^m)^{-1/2}(\thetab^\star-\hat\theta_n) &= \frac{\sqrt{m}}{\sqrt{\phi(\gamma)}}(\mathbb{V}^m)^{-1/2}(1-\gamma)^b(\theta_0-\hat\theta_n)\\ &\quad - \gamma \sum_{j=0}^{b-1} (1-\gamma)^j \frac{\sqrt{m}}{\sqrt{\phi(\gamma)}}(\mathbb{V}^m)^{-1/2}[H_n(\hat\theta_n)]^{-1} G_m^{(b-j)}(\hat\theta_n).
\end{align*}
Since the $[H_n(\hat\theta_n)]^{-1} G_m^{(b-j)}(\hat\theta_n)$ are independent (conditional on the data) and identically distributed, we have by a convolution argument:
\begin{align*}
      &\mathbb{E}^\star \left( \exp( \mathbf{i}\tau^\prime \sqrt{m}\gamma \sum_{j=0}^{b-1} (1-\gamma)^j \frac{\sqrt{m}}{\sqrt{\phi(\gamma)}}(\mathbb{V}^m)^{-1/2}[H_n(\hat\theta_n)]^{-1} G_m^{(b-j)}(\hat\theta_n) ) \right)\\ &= \prod_{j=0}^{b-1} \mathbb{E}^\star \left( \exp( \mathbf{i}\tau^\prime \sqrt{m}\gamma (1-\gamma)^j \frac{\sqrt{m}}{\sqrt{\phi(\gamma)}}(\mathbb{V}^m)^{-1/2}[H_n(\hat\theta_n)]^{-1} G_m^{(b-j)}(\hat\theta_n) ) \right)\\
      &= \prod_{j=0}^{b-1} \left[ \exp \left( -\frac{\|\tau\|_2^2}{2} \frac{\gamma^2 (1-\gamma)^{2j}}{\phi(\gamma)} \right) \left( 1+ \frac{r_m( \gamma(1-\gamma)^j \tau/\phi(\gamma) )}{m^\beta} \right) \right]\\
      &= \underbrace{ \vphantom{ \prod_{j=0}^{b-1} \left[  \left( 1+ \frac{r_m( \gamma(1-\gamma)^j \tau/\phi(\gamma) )}{m^\beta} \right) \right] } \exp \left( -\frac{\|\tau\|_2^2}{2} \frac{\gamma^2 [1-(1-\gamma)^{2b}]}{[1-(1-\gamma)^{2}]\phi(\gamma)} \right)}_{=\exp(-\|\tau\|_2^2/2)(1+o(1))} \underbrace{\prod_{j=0}^{b-1} \left[  \left( 1+ \frac{r_m( \gamma(1-\gamma)^j \tau/\phi(\gamma) )}{m^\beta} \right) \right]}_{(I)}.
\end{align*}
 To show that the last product is convergent under the stated assumptions, take logs and use the inequality $\frac{x}{1+x} \leq \log(1+x) \leq x$ for $x > -1$. Then
\begin{align*}
      \log \left(\|I\| \right) &= \sum_{j=0}^{b-1} \log \left( 1+ \frac{|r_m( \gamma(1-\gamma)^j \tau/\phi(\gamma) )|}{m^\beta} \right)
      \leq \sum_{j=0}^{b-1} \frac{|r_m( \gamma(1-\gamma)^j \tau/\phi(\gamma) )|}{m^\beta}\\
      &\leq \sum_{j=0}^{b-1} \frac{ \|\gamma \tau/\phi(\gamma)\|^\alpha (1-\gamma)^{\alpha j}}{m^\beta}
      \leq \frac{ \|\gamma \tau/\phi(\gamma)\|^\alpha }{[1-(1-\gamma)^{\alpha}] m^\beta}.
\end{align*}
Note that $\frac{\gamma}{\phi(\gamma)} = 2-\gamma \geq 1$ for $\gamma \in (0,1]$.
Putting everything together we have:
\begin{align*}
      &\mathbb{E}^\star \left( \exp( \mathbf{i}\tau^\prime \frac{\sqrt{m}}{\sqrt{\phi(\gamma)}}(\mathbb{V}^m)^{-1/2}(\thetab-\hat\theta_n) \right) = \exp\left(-\frac{\|\tau\|_2^2}{2} \right)\left( 1+O\left(\frac{ \|\tau\|^\alpha }{m^\beta} \frac{(2-\gamma)^{\alpha}}{[1-(1-\gamma)^{\alpha}]} \right)\right),
\end{align*}
which implies the desired convergence in distribution.\qed
\begin{titlingpage} 
      \emptythanks
      \title{ {Supplement to\\ \lQ {\bf Inference by Stochastic Optimization:\\ A Free-Lunch Bootstrap}''}}
      \author{Jean-Jacques Forneron\thanks{Department of Economics, Boston University, 270 Bay State Rd, MA 02215 Email: jjmf@bu.edu}  \and Serena Ng\thanks{Department of Economics, Columbia University and NBER, 420 W. 118 St. MC 3308, New York, NY 10027 Email: serena.ng@columbia.edu}}
      \setcounter{footnote}{0}
      \setcounter{page}{0}

      \clearpage 
      \maketitle 
      \thispagestyle{empty} 
      \begin{center}
      This Supplemental Material consists of Appendices \ref{apx:additional} and \ref{apx:smd} to the main text.
      \end{center}
\end{titlingpage}

\setcounter{page}{1}
\renewcommand*{\thepage}{B-\arabic{page}}

\section{Implementing rNR in R} \label{apx:Rcode}

To illustrate how the \rnr\; is implemented in a real-data setting, we provide some detailed commented \textsc{r} code below which estimates a probit model on the \citet{mroz:87} data.

\begin{lstlisting}[language=R]
      
set.seed(123)     # set the seed
library(numDeriv) # compute numerical derivaties using finite differences, alternative: library(pracma) is usually faster
library(foreign)  # to load the data set in Stata dta format


data = read.dta('mroz.dta') # read the mroz data

y = data$inlf # outcome variable
X = cbind(data$nwifeinc,data$educ,data$exper, # regressors
            data$exper^2,data$age,data$kidslt6,data$kidsge6,1) 

colnames(X) = c('nwifeinc','educ','exper','exper2', # labels
                  'age','kidslt6','kidsge6','constant')

n = 753      # sample size
index0 = 1:n # indices for the sample data

loglik <- function(coef,index=index0) {
      # compute the log-likelihood for the Probit model on the observations indexed by index (default 1:n, the original sample) at theta = coef
      
      score = X[index,]%*%coef # compute the z-scores
      ll    = y[index]*log( pnorm(score) ) +
              (1-y[index])*log( 1-pnorm(score) ) 
      return( sum( ll ) )
}

d_loglik <- function(coef,index=index0) {
      # compute the gradient of the log-likelihood for the Probit model on the observations indexed by index (default 1:n, the original sample) at theta = coef
      # In this example, the gradient is analytically tractable, it could be evaluated by finite differences by using the following:
      # d_loglik <- function(coef,index=index0) { return(jacobian(loglik,coef,index=index)) }

      yy = y[index] # keep observations indexed by index
      XX = X[index,] # keep observations indexed by index
      score = XX%*%coef # compute the z-score
      dll   = 0 # initialize the gradient
    
      for (i in 1:length(index)) {
            dll = dll +
         (yy[i]*XX[i,]*dnorm(score[i])/pnorm(score[i]) - 
         (1-yy[i])*XX[i,]*dnorm(score[i])/(1-pnorm(score[i])))
      }
      return(dll)
}


rNR <- function(coef0, learn = 0.1, iter = 500, m = n) {
      # generate 'B = iter' rNR draws with learning rate 'gamma = learn' with m out of n resampling
      
      coefs     = matrix(NA,iter,length(coef0)) # matrix where draws will be stored
      coefs[1,] = coef0 # initialize the first-draw
    
      for (i in 2:iter) {
            index = sample(1:n,m,replace=TRUE) # sample m out of n observations with replacement
            
            G = d_loglik(coefs[i-1,],index=index) # compute the resampled gradient G using analytical derivatives. Alternative using finite differences: 
            # G = jacobian(loglik,coefs[i-1,],index=index)
            H = hessian(loglik,coefs[i-1,],index=index) # compute the resampled hessian H using finite differences; we could also compute the jacobian of the gradient d_loglik
        
            coefs[i,] = coefs[i-1,] - learn*solve(H,G) # update
      }
      colnames(coefs) = colnames(X) # label the coefficients
      return( list(coefs = coefs) ) # return draws
}
# estimates and standard errors (source: Introductory Econometrics, A Modern Approach 2nd Edition, Wooldridge)
coef = c(-0.012,0.131,0.123,-0.0019,-0.053,-0.868,0.036,0.270)
ses  = c( 0.005,0.025,0.019, 0.0006, 0.008, 0.119,0.043,0.509)

iter_rNR = 2e3         # number of rNR draws
learn    = 0.3         # learning rate
coef0    = coef*3.25   # starting value

m1 = 753 # m = n
m2 = 200 # m = 200
m3 = 100 # m = 100 

# adjustments to get valid standard errors
adj_rnr1 = sqrt(m1/n)*sqrt( (1-(1-learn)^2)/learn^2 )
adj_rnr2 = sqrt(m2/n)*sqrt( (1-(1-learn)^2)/learn^2 )
adj_rnr3 = sqrt(m3/n)*sqrt( (1-(1-learn)^2)/learn^2 )

b1 = 1 + round(log(0.01)/log(1-learn)) # burn-in sample size

# generate rNR draws
out_rNR1   = rNR(coef0,learn,b1 + iter_rNR, m1)
out_rNR2   = rNR(coef0,learn,b1 + iter_rNR, m2)
out_rNR3   = rNR(coef0,learn,b1 + iter_rNR, m3)

# format output
estimates = 
      rbind( coef,
        apply(out_rNR1$coef[b1:(iter_rNR+b1),],2,mean),
        apply(out_rNR2$coef[b1:(iter_rNR+b1),],2,mean),
        apply(out_rNR3$coef[b1:(iter_rNR+b1),],2,mean))

std_errs  = 
      rbind( ses,
        apply(out_rNR1$coef[b1:(iter_rNR+b1),],2,sd)*adj_rnr1,
        apply(out_rNR2$coef[b1:(iter_rNR+b1),],2,sd)*adj_rnr2,
        apply(out_rNR3$coef[b1:(iter_rNR+b1),],2,sd)*adj_rnr3)


estimates = as.data.frame(estimates)
colnames(estimates) = colnames(X)
rownames(estimates) = c('MLE','rNRn','rNR200','rNR100')

std_errs = as.data.frame(std_errs)
colnames(std_errs) = colnames(X)
rownames(std_errs) = c('ase','rNRn','rNR200','rNR100')

# print results
print(round( cbind( t(estimates), t(std_errs) ), digits = 3 ))

# output printed below:
#            MLE   rNRn rNR200 rNR100   ase  rNRn rNR200 rNR100
#nwifeinc -0.012 -0.012 -0.013 -0.014 0.005 0.005  0.005  0.005
#educ      0.131  0.132  0.136  0.140 0.025 0.026  0.026  0.028
#exper     0.123  0.123  0.123  0.125 0.019 0.019  0.020  0.021
#exper2   -0.002 -0.002 -0.002 -0.002 0.001 0.001  0.001  0.001
#age      -0.053 -0.053 -0.054 -0.055 0.008 0.008  0.009  0.009
#kidslt6  -0.868 -0.872 -0.895 -0.917 0.119 0.121  0.121  0.126
#kidsge6   0.036  0.038  0.040  0.038 0.043 0.045  0.047  0.049
#constant  0.270  0.272  0.282  0.276 0.509 0.506  0.505  0.535
\end{lstlisting}


\newpage


\section{Additional Empirical and Simulation Results} \label{apx:additional}
 \setcounter{page}{1}
 \setcounter{table}{0}
\renewcommand*{\thepage}{C-\arabic{page}}
    

\subsection{Simulated Examples}
\paragraph{Example 2: MA(1)} The following provides additional details on computing the estimates found in Table \ref{tbl:table-MA1}. The data generating process is $y_t = \mu + e_t + \psi e_{t-1}$ where $e_t \sim \mathcal{N}(0,1)$ iid. For a given $\theta = (\mu,\psi)$, the filtered residuals are computed as $e_t(\theta) = y_t - \mu - \psi e_{t-1}(\theta)$ initialized with $e_0=0$. The \textsc{nlls} objective is then $Q_n(\theta) = \sum_{t=1}^n e_t(\theta)^2$. To find the gradient of $Q_n$ we compute the jacobian and the Hessian of $x_t(\theta) = \mu + \psi e_{t-1}(\theta)$ which are given by:
\begin{eqnarray*}
      \nabla x_t (\theta)
      = \begin{pmatrix} 1
      \\ \psi\frac{d e_{t-1}(\theta)}{d\psi}+ e_{t-1}(\theta).\end{pmatrix}, \quad 
      \nabla^2 x_t (\theta)  =  \left( \begin{array}{cc} 0 & 0 \\0 & \frac{d e_{t-1}(\theta)}{d\psi} \end{array} \right).
\end{eqnarray*}
The gradient of $Q_n$ is $G_n(\theta)=2\sum_{t=1}^n e_t(\hat \theta_n)\nabla x_t(\hat\theta_n)=0$. Similarly, the Hessian is $H_n(\theta)=2\sum_{t=1}^n [ e_t(\hat \theta_n)\nabla^2 x_t(\hat\theta_n) + \nabla x_t(\hat\theta_n) \nabla^\prime x_t(\hat\theta_n)]$. The objective is minimized using Newton-Raphson iterations based on the analytical $G_n,H_n$. The asymptotic standard errors are computed from the inverse Hessian, based on the information matrix equality. 

For the standard bootstrap, we implement a resampling scheme desgined for State-Space models described in \citet{stoffer2004}. Given a converged estimate $\hat\theta_n$, compute the filtered $e_t(\hat\theta_n)$. The resampled data is then generated as $y_t^{(b)} = \hat \mu_n +  e^{(b)}_t(\hat\theta_n) + \hat \psi_n e^{(b)}_{t-1}(\hat\theta_n)$ where $e^{(b)}_t(\hat\theta_n)$ are iid draws with replacement taken from $\{ \hat e_t(\hat\theta_n) \}_{t=1,\dots,n}$. The resampled \textsc{nlls} objective $Q_n^{(b)}(\theta)$ is then computed and minimized as described above. This procedure is very time-consuming and is implemented in C++ using Rcpp to reduce computation time. Other methods described below are implemented using only \textsc{r}. 

To implement \dmk, given a converged estimate $\hat\theta_n$, filtered residuals $e_t (\hat\theta_n)$ and their derivates, we sample indices $t_{1,b},\dots,t_{n,b}$ with replacement from $\{1,\dots,n\}$ for each $b$ and compute the resampled gradient and Hessian as $G_n^{(b)} =2\sum_{j=1}^n e_{t_{j,b}}(\hat \theta_n)\nabla x_{t_{j,b}}(\hat\theta_n)$ and $H_n^{(b)} =2\sum_{j=1}^n [e_{t_{j,b}}(\hat \theta_n)\nabla^2 x_{t_{j,b}}(\hat\theta_n) + \nabla x_{t_{j,b}}(\hat\theta_n) \nabla^\prime x_{t_{j,b}}(\hat\theta_n)]$. We then generate the draws using one \nr\; iteration $\theta_{\dmk}^{(b)} = \hat\theta_n - [H_n^{(b)}(\hat\theta_n)]^{-1} G_n^{(b)}(\hat\theta_n)$.

To implement \rnr\, with $m \leq n$, sample a block of $m$ observations $(y_1^{(b)},\dots,y_{m}^{(b)})=(y_{t},y_{t+1},\dots,y_{t+m})$ with $1 \leq t \leq n-m+1$ and compute the filtered residuals $e^{(b)}_t(\theta_{b-1}) = y_t^{(b)} - \mu_{b-1} - \psi_{b-1} e^{(b)}_{t-1}(\theta_{b-1})$ for $t=1,\dots,m$ where $(\mu_{b-1},\psi_{b-1}) = \theta_{b-1}$ is the previous \rnr\, draw. As above, the filtered residuals are initialized at $e_0=0$ and the \rnr\, draws are initialized at $\theta_0 = (0,0)$. Similarly to our implementation of \dmk, we then we sample indices $t_{1,b},\dots,t_{m,b}$ with replacement from $\{1,\dots,m\}$ and compute the resampled gradient and Hessian $G_m^{(b)},H_m^{(b)}$, the updating equation gives the draws $\theta_b = \theta_{b-1} - \gamma [H_m^{(b)}(\theta_{b-1})]^{-1} G_m^{(b)}(\theta_{b-1})$.

\paragraph{Size of Confidence Intervals in the Simulated Examples}
The table below presents the size of confidence intervals over $1000$ replications in the simulated examples of Section \ref{sec:examples}. Frequentist confidence intervals (\textsc{ase}) are computed using $\hat\theta_n \pm 1.96 \text{se}(\hat\theta_n)$. Bootstrap confidence intervals for the standard bootstrap (\textsc{boot}), \dmk\; and \ks\; are computed by taking the $2.5$ and $97.5\%$ percentiles of the draws $\theta^{(b)}$ except for the dynamic panel as discussed below. For \rnr, the confidence intervals are computed by taking the $2.5$ and $97.5\%$ percentiles of $\overline{\theta}_{\FL}+\sqrt{\frac{m}{n\phi(\gamma)}}(\theta_b - \overline{\theta}_{\FL})$ where $\phi(\gamma) = \frac{\gamma^2}{1-(1-\gamma)^2}$.

For the dynamic panel, the standard bootstrap (\textsc{boot}), \dmk\; and \ks\; draws are adjusted so that confidence intervals are computed by taking the $2.5$ and $97.5\%$ percentiles of $\hat\theta_{n,\SMD} + (\theta^{(b)} - \overline{\theta}_B)$, where $\overline{\theta}_B$ is the average bootstrap draw. Without this recentering the confidence intervals display significant size distortion, see Appendix \ref{apx:smd} for a discussion of this recentering. For \rnr, we take the $2.5$ and $97.5\%$ percentiles of $\overline{\theta}_{\FL,S}+\sqrt{\frac{m}{n\phi(\gamma)}}\theta_{b,S}^2$ where $\overline{\theta}_{\FL,S} = \frac{1}{B} \sum_{b=1}^B \theta^1_{b,S}$ after discarding the burn-in draws. 

\subsection{Empirical Examples}
\paragraph{Application 1: Labor Force Participation}
The table below presents the estimates and standard errors for all methods and coefficients in the \citet{mroz:87} application.

\begin{table}[H]
      \centering  \caption{Labor Force Participation: Estimates and Standard Errors} \label{tbl:table-mroz_apx} \setlength\tabcolsep{4.5pt}
      { \renewcommand{\arraystretch}{0.935} 
      \begin{tabular}{l|bbbbaaaaaa}
            \hline \hline 
            & \multicolumn{10}{c}{Estimates}\\
       & \textsc{mle}  &  &  &  &  \mc{1}{\rnr$_n$} & \mc{1}{\rnr$_{200}$} & \mc{1}{\rnr$_{100}$} & \mc{1}{r\textsc{qn}$_n$} & \mc{1}{r\textsc{qn}$_{200}$} & \multicolumn{1}{c}{r\textsc{qn}$_{100}$}  \\ 
        \hline
        nwifeinc & -0.012 & - & - & - & -0.012 & -0.013 & -0.014 & -0.012 & -0.011 & -0.012 \\ 
  educ & 0.131 & - & - & - & 0.132 & 0.138 & 0.143 & 0.131 & 0.129 & 0.129 \\ 
  exper & 0.123 & - & - & - & 0.123 & 0.124 & 0.123 & 0.123 & 0.124 & 0.125 \\ 
  exper2 & -0.002 & - & - & - & -0.002 & -0.002 & -0.002 & -0.002 & -0.002 & -0.002 \\ 
  age & -0.053 & - & - & - & -0.053 & -0.053 & -0.055 & -0.052 & -0.052 & -0.052 \\ 
  kidslt6 & -0.868 & - & - & - & -0.874 & -0.892 & -0.902 & -0.864 & -0.855 & -0.844 \\ 
  kidsge6 & 0.036 & - & - & - & 0.037 & 0.038 & 0.041 & 0.036 & 0.035 & 0.032 \\ 
  const. & 0.270 & - & - & - & 0.271 & 0.216 & 0.234 & 0.248 & 0.256 & 0.249 \\
         \hline
         & \multicolumn{10}{c}{Standard Errors} \\
         & \textsc{ase} & \textsc{boot} & \dmk & \ks & \mc{1}{\rnr$_n$} & \mc{1}{\rnr$_{200}$} & \mc{1}{\rnr$_{100}$} & \mc{1}{r\textsc{qn}$_n$} & \mc{1}{r\textsc{qn}$_{200}$} & \multicolumn{1}{c}{r\textsc{qn}$_{100}$}\\  \hline
         nwifeinc & 0.005 & 0.005 & 0.005 & 0.005 & 0.005 & 0.006 & 0.005 & 0.005 & 0.005 & 0.005 \\ 
  educ & 0.025 & 0.026 & 0.026 & 0.025 & 0.025 & 0.027 & 0.028 & 0.027 & 0.025 & 0.025 \\ 
  exper & 0.019 & 0.020 & 0.019 & 0.019 & 0.019 & 0.020 & 0.021 & 0.019 & 0.018 & 0.017 \\ 
  exper2 & 0.001 & 0.001 & 0.001 & 0.001 & 0.001 & 0.001 & 0.001 & 0.001 & 0.001 & 0.001 \\ 
  age & 0.008 & 0.009 & 0.008 & 0.008 & 0.009 & 0.008 & 0.009 & 0.009 & 0.008 & 0.008 \\ 
  kidslt6 & 0.119 & 0.120 & 0.118 & 0.118 & 0.120 & 0.119 & 0.129 & 0.117 & 0.113 & 0.117 \\ 
  kidsge6 & 0.043 & 0.046 & 0.045 & 0.045 & 0.045 & 0.048 & 0.047 & 0.044 & 0.042 & 0.045 \\ 
  const. & 0.509 & 0.512 & 0.507 & 0.505 & 0.494 & 0.535 & 0.544 & 0.544 & 0.494 & 0.506 \\ \hline  \hline
      \end{tabular} }
\end{table}

We also implemented \rgd\; in this application to evaluate its feasibility in a real-data setting. We found that \rgd\, requires a burn-in greater than $1000$ draws to converge and the high persistence of the draws results in a very small effective sample size while the \rnr\, converges quickly ($\leq 20$ draws) and has good mixing properties. This is mainly due to the ill-conditioning of the problem since $\frac{\lambda_\min(H_n)}{\lambda_\max(H_n)}$  evaluated at $\theta=\hat\theta_n$ is $10^{-7}$ which implies a very slow convergence for \gd\, \sgd\; and \rgd.

\paragraph{Application 2: Earnings Dynamics}
During the initial convergence phase some adjustments to the \rnr\, updating equations were required to handle the non-convexity of the objective in the \citet{moffitt-zhang:18} application. For \rnr\; and r\textsc{qn}, draws such that the sample objective $Q_n$ increases $6$-folds or more are discarded, i.e. we only keep $\theta_b$ if $Q_n(\theta_b) \leq 6 Q_n(\theta_{b-1})$. This never occurs for \rnr\, and \rnr$_w$. It occurred twice for r\textsc{qn} and four times for r\textsc{qn}$_w$  but only in the burn-in sample with $\textsc{burn}=50$. When a draw is discarded, the \textsc{bfgs} approximation of the Hessian is reset to the Hessian computed using finite differences. These adjustments ensured that r\textsc{qn} converged from the original starting values. For \rnr$_{w}$, we reweight the observations using exponential $\mathcal{E}(1)$ draws. For \ks, the score is reweighted using draws from the Rademacher distribution. Results presented in Table \ref{tbl:table-earnings} were computed using $4$ cluster nodes with an eight-core 2.6 GHz Intel Xeon E5-2650v2 processor.

\paragraph{Application 3: Demand for Cereal} To side-step possible identification issues, we omit the income$^2 *$price interaction as well as the child$*$price and age-related coefficients. The results are broadly similar when the child$*$price coefficient is included.

The \textsc{r} package \textsc{BLPestimatoR} does not offer a bootstrap option. The `parametric' bootstrap  implemented in the Python \textsc{pyBLP} package of \citet{conlon2019}  draws from the asymptotic distribution, making Gaussian draws centered at $\hat\theta_n$ with a sandwich variance-covariance matrix. \textsc{BLPestimatoR}, implements estimation taking as input a dataset, initial values and a model specification. To implement the standard bootstrap using this package, we simply update the data by resampling at the market level and use the built-in functions to re-estimate using as initial value the sample estimate $\hat \theta_n$. For \dmk,  \rnr\; and r\textsc{qn} the data is updated as described above, then built-in functions provide analytical gradient estimates. The Hessian is computed for \dmk\; and \rnr\; using finite differences. 

Table \ref{tab:blp_gammas} replicates the \rnr\; estimates and standard errors from Section \ref{sec:examples} with different learning rates $\gamma=0.1,0.3,0.6$. The results are similar using $\gamma \in [0.1,0.3]$ while $\gamma=0.6$ is less stable and results in large standard errors for the income$*$price interaction coefficient.

\begin{table}[ht] \caption{Demand for Cereal: Estimates and Standard Errors for $\gamma = 0.1,0.3,0.6$} \label{tab:blp_gammas}
      \centering
      { \renewcommand{\arraystretch}{0.935} 
      \begin{tabular}{cl|baaa|baaa}
            \hline \hline 
            & & \multicolumn{4}{c|}{Estimates} & \multicolumn{3}{c}{Standard Errors} \\ \hline
            & & $\hat\theta_n$ & \mc{1}{\rnr$_{0.1}$} & \mc{1}{\rnr$_{0.3}$} & \multicolumn{1}{c|}{\rnr$_{0.6}$} & \textsc{boot} & \mc{1}{\rnr$_{0.1}$} & \mc{1}{\rnr$_{0.3}$} & \mc{1}{\rnr$_{0.6}$} \\
            \hline \parbox[t]{2mm}{\multirow{4}{*}{\rotatebox[origin=c]{90}{stdev}}}
            & const. & 0.284 & 0.264 & 0.266 & 0.260 & 0.129 & 0.126 & 0.127 & 0.176 \\ 
            & price & 2.032 & 2.191 & 2.183 & 2.162 & 1.198 & 0.930 & 1.013 & 1.689 \\ 
            & sugar & -0.008 & -0.006 & -0.006 & -0.005 & 0.017 & 0.011 & 0.011 & 0.017 \\ 
            & mushy& -0.077 & -0.057 & -0.056 & -0.057 & 0.177 & 0.151 & 0.163 & 0.233 \\ \hline \parbox[t]{2mm}{\multirow{4}{*}{\rotatebox[origin=c]{90}{income}}}
            & const. & 3.581 & 3.475 & 3.463 & 3.459 & 0.666 & 0.721 & 0.747 & 1.451 \\ 
            & price & 0.467 & 1.235 & 1.360 & 1.255 & 3.829 & 3.744 & 4.187 & 16.458 \\ 
            & sugar & -0.172 & -0.170 & -0.170 & -0.166 & 0.028 & 0.029 & 0.028 & 0.135 \\ 
            & mushy & 0.690 & 0.643 & 0.634 & 0.535 & 0.345 & 0.353 & 0.355 & 1.582 \\ 
         \hline \hline
      \end{tabular}
      }
      \end{table}

Results presented in Table \ref{tbl:table-earnings} were computed using $4$ cluster nodes with a fourteen-core 2.4 GHz Intel Xeon E5-2680v4 processor. The CH estimates were computed on a different batch job and were assigned at runtime to an eight-core 2.6 GHz Intel Xeon E5-2670 processor.

\clearpage
\setcounter{page}{1}
\setcounter{table}{0}
\renewcommand*{\thepage}{D-\arabic{page}}

\section{SMD Estimation} \label{apx:smd}

\begin{table}[ht]
      \centering \caption{Dynamic Panel: Estimates of $\rho$ and Standard Errors} \label{tab:panel_apx}
{ 
\begin{tabular}{ll|baaa|bbbaaa}
      \hline \hline
     & &  \multicolumn{4}{c|}{Estimates}  &  \multicolumn{6}{c}{Standard Errors} \\ 
   $S$ &  $m$  & \mc{1}{\textsc{ind}} & \mc{1}{\rnr$_{0.3}$} & \mc{1}{\rnr$_{0.1}$} & \multicolumn{1}{c|}{\rnr$_{0.01}$} & \textsc{ase} & \textsc{boot} & \dmk & \mc{1}{\rnr$_{0.3}$} & \mc{1}{\rnr$_{0.1}$} & \mc{1}{\rnr$_{0.01}$}\\
        \hline
        & $200$ & 0.619 & 0.589 & 0.592 & 0.590 & 0.045 & 0.049 & 0.050 & 0.034 & 0.034 & 0.025 \\ 
       1 & $100$ & - & 0.589 & 0.588 & 0.586 & - & 0.048 & - & 0.036 & 0.039 & 0.037 \\ 
        & $50$ & - & 0.580 & 0.588 & 0.581 & - & 0.050 & - & 0.037 & 0.037 & 0.024 \\  \hline
        & $200$ & 0.604 & 0.587 & 0.589 & 0.588 & 0.041 & 0.042 & 0.041 & 0.036 & 0.034 & 0.024 \\ 
       2 & $100$ & - & 0.588 & 0.587 & 0.589 & - & 0.041 & - & 0.033 & 0.036 & 0.037 \\ 
        & $50$ & - & 0.588 & 0.592 & 0.581 & - & 0.041 & - & 0.037 & 0.038 & 0.038 \\  \hline
        & $200$ & 0.578 & 0.589 & 0.590 & 0.590 & 0.037 & 0.038 & 0.037 & 0.033 & 0.035 & 0.034 \\ 
       5 & $100$ & - & 0.591 & 0.589 & 0.588 & - & 0.038 & - & 0.036 & 0.038 & 0.037 \\ 
        & $50$ & - & 0.583 & 0.581 & 0.581 & - & 0.038 & - & 0.037 & 0.035 & 0.029 \\  \hline
        & $200$ & 0.584 & 0.591 & 0.589 & 0.589 & 0.035 & 0.036 & 0.036 & 0.035 & 0.036 & 0.032 \\ 
       10 & $100$ & - & 0.589 & 0.591 & 0.587 & - & 0.034 & - & 0.034 & 0.037 & 0.030 \\ 
        & $50$ & - & 0.586 & 0.589 & 0.587 & - & 0.035 & - & 0.038 & 0.031 & 0.032 \\ 
         \hline \hline
      \end{tabular}
      }\\
      \textit{Remark: Results reported for one simulated sample of size $n=200, T=5$.}
\end{table}

 \subsection*{SMD Estimation of a Sample Mean} \label{apx:smd_mean}
To illustrate Proposition \ref{cor:smd_rnr}, consider the simple model $y_i \sim \mathcal{N}(\theta^0,1)$. The \textsc{md} estimator of $\theta^0$ is $\hat\theta_{\textsc{md}}=\overline{y}_n\equiv \hat\psi_n$. For any given $\theta$, let $y_i^s(\theta) = \theta + e_i^s$ where $e_i^s \sim \mathcal{N}(0,1)$.  The \textsc{smd} estimator is the $\theta$ that equates  $\psi(\theta,y_S)=\overline{y}_{n,S}(\theta)$ to  $\hat\psi_n$, and is found to be  $\hat\theta_{\SMD} = \hat\theta_{\MD} - \overline{e}_{n,S}$. The \rnr\; resamples and simulates the binding function to give
\[ \theta^1_{b+1,S} - \hat\theta_{n,\MD} = (1-\gamma)(\theta^1_{b,S} - \hat\theta_{n,\MD}) + \gamma (\hat\theta_{m,\MD}^{(b)}-\hat\theta_{n,\MD} - \overline{e}_{m,S}^{(b)}). \]
Note that resampling alone gives $\thetabf-\hat\theta_{n,\MD}=(1-\gamma)(\thetab-\hat\theta_{n,\MD})+\gamma (\hat\theta_{m,\MD}^{(b)}-\hat\theta_{n,\MD})$.  Taking conditional expectations, we have $\mathbb{E}^\star(\theta^1_{b+1,S}) = \hat\theta_{n,\MD} + (1-\gamma)^{b+1}(\theta_0-\hat\theta_{n,\MD})$ so that $\mathbb{E}^\star(\overline{\theta}_{\FL,S}) = \hat\theta_{n,\MD} + O(\frac{1}{B})$, as in the OLS example. Furthermore, $\text{var}^\star(\overline{\theta}_{\FL,S}) = O(\frac{1}{mB} + \frac{1}{mSB})$ where the first term is due to resampling ($\hat\theta_{m}^{(b)}-\hat\theta_{n,\MD}$) and the second is due to simulation noise ($\overline{e}_{m,S}^{(b)}$). Hence for this example, $\overline{\theta}_{\FL} = \hat\theta_{n,\MD} + O_{p^\star}(\frac{1}{B} + \frac{1}{\sqrt{mB}} + \frac{1}{\sqrt{mSB}} )$, showing that by averaging over both the resampling and simulation noise, $\overline\theta_{\FL,S}$  is first-order equivalent to $\hat\theta_{n,\MD}$  if $\frac{\sqrt{n}}{\min(m,B)} \to 0$ for any $S \geq 1$. 

Part ii. of the Proposition involves a second sequence $\theta^2_{b,S}$ because
the variance of the \rnr\, draws are comprised of  two quantities: $\text{var}^\star(\hat\theta_m^{(b)})$ and $\text{var}^\star(e_{m,S}^{(b)})$. But   $\frac{m}{\phi(\gamma)}\text{var}^\star(\thetab) = m \text{var}^\star(\hat\theta_m^{(b)}) + m \text{var}^\star(e_{m,S}^{(b)}) > m \text{var}^\star(\hat\theta_m^{(b)})$, and as a consequence $\var^*(\hat\theta_m^{(b)})$ is larger than the actual sampling uncertainty of $\overline{\theta}_{\FL,S}$. Running the second chain  in parallel using the same resampled statistic $\{ \overline{y}_{m}^{(b)} \}_{b=1,\dots,B}$  produces the AR(1) draws $\{\theta_{b,S}^2\}$. We can rewrite  these draws as:
\[ \theta^2_{b+1,S} = (1-\gamma)\theta^2_{b,S} + \gamma (\hat \theta^{(b)}_{m,\textsc{md}}-\hat \theta_{n,\textsc{md}}).\]
This is  an AR(1) process that targets the infeasible sampling distribution based on the intractable \textsc{md} objective function. 
From the OLS example, we know that $\text{var}^\star ( \theta^2_{b,S}) = \frac{\gamma^2 + o(1)}{1-[1-\gamma]^2} \text{var}^\star ( \hat\theta^{(b)}_{m,\textsc{md}})$ which is proportional to the desired variance. Hence, $\mathbb V_{\FL,S}=\frac{m}{\phi(\gamma)}\text{var}^\star(\thetab^2) = m \text{var}^\star(\hat\theta_{m,\textsc{md}}^{(b)})$ yields valid standard errors for $\overline{\theta}_{\FL,S}$.

Note also that  the \textsc{smd} bootstrap draws $\hat\theta^{(b)}_{m,\textsc{smd}} = \hat\theta^{(b)}_{m,\textsc{md}} - \overline{e}_{m,S}^{b}$ are centered around $\hat\theta_{n,\textsc{md}}$ instead of $\hat\theta_{n,\textsc{smd}}$ because the simulation noise $\overline{e}_{m,S}^{b}$ averages out. Since  $\mathbb{E}^\star (\hat\theta^{(b)}_{m,\textsc{smd}}) = \hat\theta_{n,\textsc{md}}$, 
the bootstrap confidence interval must be re-centered around $\hat\theta_{n,\textsc{smd}}$ to have correct size. In contrast with \rnr, the variance does not need to be adjusted. In the numerical examples below, the draws were re-centered around $\hat\theta_{n,\textsc{smd}}$. A numerical illustration of this example is given below.

\paragraph{Example 4: Sample Mean} 
To illustrate that the \rnr\; draws achieve the same efficiency as an \textsc{smd} estimators with $S=\infty$ at a lower computation cost of  $S=1$, we simulate  $y_i \sim \mathcal{N}(\theta,1)$ with $\theta=1$, $n=1000$. 
Table \ref{tab:mean_variance} illustrates the variance properties of $\rnr\,$ relative to indirect inference and the size of confidence intervals derived from the quantiles of the draws $\theta^2_{b,S}$. With $m=200 < n=1000$, the variance of $\rnr$\, is comparable to the method of moments (which has no simulation noise) and indirect inference with $S=20$ simulated samples of $n=1000$ observations. The size of $m$ out of $n$ bootstrap confidence intervals are reported in the last line of the table for each estimator. Size for \rnr\, is again comparable to the method of moments and indirect inference. 

\begin{table}[ht] \caption{Mean Estimaton: standard deviation and size}
      \centering \label{tab:mean_variance}
      \begin{tabular}{l|barrrr}
        \hline\hline
       & \textsc{mm} & \mc{1}{\rnr} & \textsc{ind}$_1$ & \textsc{ind}$_5$ & \textsc{ind}$_{10}$ & \textsc{ind}$_{20}$ \\
        \hline
std & 0.031 & 0.032 & 0.047 & 0.035 & 0.033 & 0.032 \\ 
size & 0.059 & 0.056 & 0.059 & 0.059 & 0.044 & 0.050 \\ 

         \hline\hline
      \end{tabular}\\
      \textit{Legend: $n=1000$; \rnr\, $\gamma=0.3, m=200, B=1000$;\\ \textsc{ind}$_S$: indirect inference with $S=1,5,10,20$. $1000$ replications.}
\end{table}

\newpage
\end{appendices}
\end{document}